\documentclass{aa}
\usepackage[varg]{txfonts}
\usepackage{graphicx}				
\usepackage{xcolor}
\usepackage{subfig}
\usepackage{import}
\usepackage[percent]{overpic}
\usepackage{textcomp}
\usepackage{url}

\usepackage{ulem, cancel} 

\newcommand{\pyeta}{$^\pi \eta$ }

\usepackage{verbatim}

\newcommand{\centralbibliography}{/Users/pino/Documents/bibliography}
\bibpunct{(}{)}{;}{a}{}{,}

\begin{document}
\title{Combining low- to high-resolution transit spectroscopy of HD189733b}
\subtitle{Linking the troposphere and the thermosphere of a hot gas giant}
\author{Lorenzo Pino \inst{1,2}
\and David Ehrenreich \inst{2}
\and Aur\'elien Wyttenbach \inst{2}
\and Vincent Bourrier \inst{2}
\and Valerio Nascimbeni \inst{1}
\and Kevin Heng \inst{3}
\and Simon Grimm\inst{3}
\and Christophe Lovis \inst{2}
\and Matej Malik \inst{3}
\and Francesco Pepe \inst{2}
\and Giampaolo Piotto \inst{1}}

\institute{Dipartimento di Fisica e Astronomia `Galileo Galilei', Univ. di Padova, Vicolo dell’Osservatorio 3, Padova I-35122, Italy
\and Observatoire astronomique de l’Universit\'e de Gen\`eve, Universit\'e de Gen\`eve, 51 chemin des Maillettes, CH-1290 Versoix, Switzerland
\and University of Bern, Center for Space and Habitability, Sidlerstrasse 5, CH-3012, Bern, Switzerland}

				
\abstract{Space-borne low- to medium-resolution ($\mathcal{R}\sim10^{2}-10^{3}$) and ground-based high-resolution spectrographs ($\mathcal{R}\sim10^5$) are commonly used to obtain optical and near infrared transmission spectra of exoplanetary atmospheres. In this wavelength range, space-borne observations detect the broadest spectral features (alkali doublets, molecular bands, scattering, etc.), while high-resolution, ground-based observations probe the sharpest features (cores of the alkali lines, molecular lines). The two techniques differ by several aspects. (1) The Line Spread Function of ground-based observations is $\sim10^3$ times narrower than for space-borne observations; (2) Space-borne transmission spectra probe up to the base of thermosphere ($P\gtrsim10^{-6}~\mathrm{bar}$), while ground-based observations can reach lower pressures (down to $\sim10^{-11}~\mathrm{bar}$) thanks to their high resolution; (3) Space-borne observations directly yield the transit depth of the planet, while ground-based observations can only measure differences in the apparent size of the planet at different wavelengths. These differences make it challenging to combine both techniques. Here, we develop a robust method to compare theoretical models with observations at different resolutions. We introduce $^\pi \eta$, a line-by-line 1D radiative transfer code to compute theoretical transmission spectra over a broad wavelength range at very high resolution ($\mathcal{R}\sim10^6$, or $\Delta\lambda\sim0.01~\mathrm{\AA}$). An hybrid forward modeling/retrieval optimization scheme is devised to deal with the large computational resources required by modeling a broad wavelength range ($\sim0.3-2~\mathrm{\mu m}$) at high resolution. We apply our technique to HD189733b. In this planet, HST observations reveal a flattened spectrum due to scattering by aerosols, while high-resolution ground-based HARPS observations reveal sharp features corresponding to the cores of sodium lines. We reconcile these apparent contrasting results by building models that reproduce simultaneously both data sets, from the troposphere to the thermosphere.  We confirm: (1) the presence of scattering by tropospheric aerosols; (2) that the sodium core feature is of thermospheric origin. When we take into account the presence of aerosols, the large contrast of the core of the sodium lines measured by HARPS indicates a temperature of up to $\sim 10\,000$ K in the thermosphere, higher than what reported in the literature. We also show that the precise value of the thermospheric temperature is degenerate with the relative optical depth of sodium, controlled by its abundance, and of the aerosol deck.}
				
\maketitle				

\section{Introduction}
By studying the variation in the apparent radius of a planet as a function of wavelength, i.e. by building its transmission spectrum, it is possible to retrieve the physical conditions and composition of its atmosphere. Despite efforts to use low-resolution spectroscopy and accurate photometry to obtain transmission spectra from the ground (e.g. \citealt{Sedaghati2015,  Nascimbeni2013}), space-borne low- to medium-resolution spectroscopy ($\mathcal{R}\sim10^2-10^3$) and ground-based high-resolution spectroscopy ($\mathcal{R}\sim10^5$) remain the most solid techniques to date.\\
In the near-infrared (NIR), HST/WFC3 transmission spectra are available for tens of planets, comprising several hot Jupiters, some hot and warm Neptunes, and even some super-Earths (e.g. \citealt{Kreidberg2015, Ehrenreich2014, Knutson2014}). Observations of this kind are sensitive to water absorption bands at $1.1~\mu \mathrm{m}$ and $1.4~\mu \mathrm{m}$, whose strength is often lower than predicted by theoretical models. Both aerosols \citep{ Seager2000, Ehrenreich2012, Howe2012} and subsolar abundance of water \citep{Seager2005, Madhusudhan2014} could be responsible for weakening or even concealing these spectral features. Combining optical and infrared HST observations, \cite{Sing2016} were able to break this degeneracy; their analysis favoured the aerosol explanation (see also \citealt{Barstow2017}).\\
On the other side of the transmission spectrum, optical observations with HST/ACS and HST/STIS led to the detection of an enhanced optical slope in the transmission spectra of several exoplanets \citep{Etangs2008, Pont2008, Sing2016}, which can be explained by scattering from small aerosols (see section \ref{continuum} and e.g. \citealt{Ehrenreich2014, Sing2015}). The strongest spectral features, such as the sodium and potassium doublets, are generated above the aerosol deck and are thus visible in transmission spectra (e.g. \citealt{Charbonneau2002}, \citealt{Sing2015}).
The broad spectral coverage and the accuracy of the optical and NIR HST observations allowed us to firmly detect many species and spectral features in exoplanet atmospheres ($\mathrm{Na}$, $\mathrm{K}$, $\mathrm{H_2O}$, aerosols, etc.), but it is limited by relatively low resolving powers ($\mathcal{R}\sim10^2-10^3$, see Eq. \ref{eq:resolving_power} and Table \ref{tab:observations}).\\
Ground-based observatories are able to achieve higher resolving powers of  $\mathcal{R}\sim10^5$. At such high spectral resolution, single absorption lines are spectrally resolved and molecular fingerprints are uniquely identified \citep{Snellen2010, DeKok2014, Brogi2016, Allart2017}. Even the single lines of the sodium and potassium doublets are resolved, and it is possible to probe lower pressures at high altitude in the atmosphere through their sharp cores \citep{ Vidal-Madjar2011, Vidal-Madjar2011b, Wyttenbach2015}.\\
Space-born low-resolution and ground-based high-resolution transit spectroscopy are complementary. However, they are difficult to combine because the data-reduction process for the two techniques is different. Both techniques are time-differential, meaning that they compare stellar spectra taken at different times to record the transit radius. Ground-based, high-resolution transmission spectroscopy is `double-differential', i.e. it also requires a wavelength differentiation. The wavelength differentiation is necessary to eliminate variations in the measured flux only due to instrumental effects and the atmosphere of the Earth \citep{Snellen2008, Wyttenbach2015, Heng2015, Wyttenbach2017}. During wavelength differentiation the absolute level of the absorption of the planet is lost. Thus, assembling transmission spectra obtained by the two techniques requires a common normalization framework.\\
Another challenge is the interpretation of the transit radius in terms of models. In particular, several authors have noted that the reference radius/pressure  level and the abundance of atmospheric constituents are degenerate \citep{Etangs2008, Griffith2014, Heng2017}. However, in this paper we do not infer absolute abundances for the atmospheric constituents of the atmosphere, and the presence of this degeneracy does not affect our conclusions.\\
\cite{Brogi2017} presented a first attempt to combine high- and low-resolution transmission spectra. They combined high-resolution, ground-based, infrared ($\lambda\sim2.287-2.345~\mathrm{\mu m}$) CRIRES eclipse observations with HST WFC3 observations ($\lambda\sim1.125-1.655~\mathrm{\mu m}$) of lower resolution but more extended spectral coverage, which were used by \cite{Line2016} to retrieve the atmospheric conditions in HD209458b. The results show that high-resolution spectra on a narrow band are sufficient to increase the precision on the retrieved abundances of the molecular constituents of the atmosphere by at least one order of magnitude (several for CO).\\
Here, we combine for the first time optical high-resolution, ground-based transmission spectroscopy and optical and NIR space-borne transmission spectroscopy of one of the most well-studied hot Jupiters, HD189733b \citep{Bouchy2005}. Low- to medium-resolution data probe atmospheric layers from the troposphere ($P\sim\mathrm{10~bar-10^{-4}~bar}$), where clouds, molecular bands and the wings of the alkali doublets are generated, to the base of the thermosphere (down to $P\gtrsim 10^{-6}~\mathrm{bar}$). On the other hand, high-resolution optical data are sensitive to the fine cores of the alkali atoms \citep{Wyttenbach2015, Wyttenbach2017} generated at higher altitudes in the lower thermosphere ($P\sim\mathrm{10^{-4}~bar-10^{-11}~bar}$). These observations are complementary to near-UV transit spectroscopy, which probes the transition between the upper thermosphere and the exosphere \citep{Vidal-Madjar2013}, and to far-UV transit spectroscopy, which probes the exosphere. This is the outermost region of the atmosphere of the planet. Thermospheric heating by stellar X/EUV photons into the lower thermosphere is the source of the expansion of the HD189733b upper atmosphere \citep{Vidal-Madjar2003, Lammer2003, Yelle2004, Vidal-Madjar2011,Vidal-Madjar2011b, Koskinen2013}, which leads to the escape of hydrogen and heavier particles into the exosphere \citep{Etangs2010, Etangs2012, Bourrier2013, Ben-Jaffel2013}.\\
For HD189733b, the low-resolution transmission spectrum is flattened by the scattering of stellar light by aerosols suspended in the troposphere of the planet \citep{Pont2013, Sing2016}; instead, the ground-based, high-resolution transmission spectrum reveals the presence of sharp absorption features in the core of the sodium doublet lines \citep{Wyttenbach2015}. These results are apparently contrasting. By modelling the two data sets simultaneously we aim to reconcile them. Furthermore, \cite{Vidal-Madjar2011,Vidal-Madjar2011b} and \cite{Wyttenbach2015} have measured a positive temperature gradient through the sodium doublet, indicative of thermospheric heating (the \citealt{Wyttenbach2015} result was confirmed by \citealt{Heng2015}, via an independent analysis of the same data set). With our analysis we aim to investigate the biases and degeneracies introduced by the presence of tropospheric aerosols (neglected in previous works) in the retrieval based on high-resolution data.\\
Using the \pyeta code, which we present in section \ref{sec:Pyeta}, we produce high-resolution ($R\sim10^6$) transmission spectrum models in order to simulate HST and HARPS data sets. These simulations are described in sections \ref{compobs} and \ref{caveat}. High-resolution models with a broad spectral coverage are computationally very demanding. Instead of performing a full retrieval approach which requires the computation of up to millions of models \citep{Madhusudhan2009, Lavie2016}, we adopt a hybrid forward modelling/retrieval approach. In section \ref{sec:methods}, we describe this method. In section \ref{results}, we discuss the models that best reproduce the data and focus on their consistency with both low- and high-resolution spectra. Furthermore, we explore how the different data sets are sensitive to different atmospheric parameters and discuss their complementarity.

\section{The \pyeta code} \label{sec:Pyeta}
We introduce \pyeta (a Python code for extra-solar transiting atmospheres), an improved version of the $\eta$ code presented in \cite{Ehrenreich2006} and expanded in \cite{Ehrenreich2012} to compute transmission spectra of exoplanetary atmospheres. There are three main characteristics of \pyeta:
\begin{enumerate}
\item High-resolution ($\mathcal{R}\sim10^6$). This is necessary to compare models with ground-based, high-resolution data;
\item Broad wavelength coverage ($200~\mathrm{nm}-2~\mu\mathrm{m}$). This is necessary to compare models with space-borne, optical to NIR data;
\item Flexibility of the input composition and $T-p$ profile. \pyeta is used to model simultaneously several orders of magnitudes in pressure where common assumptions such as equilibrium chemistry may break down.
\end{enumerate}
In the following we summarize the basic concept of $\eta$ (see \citealt{Ehrenreich2006} and \citealt{Brown2001} for a more complete description of the basic equations of the problem). We then describe the improvements of \pyeta with respect to its former version.
\subsection{The $\eta$ code}
The $\eta$ code is a plane parallel, line-by-line radiative transfer code used to compute the transmission spectra $\delta(\lambda)$ of exoplanetary atmospheres. More precisely, $\delta(\lambda)$ is the transit depth, i.e. the flux absorbed by the planetary atmosphere in units of stellar flux. The total opacity arising from photoabsorption by atoms and molecules and scattering by molecules and aerosols in the atmosphere determines the spectral shape of the absorption. Practically, this code computes $\tau_{b}(\lambda)$, the total optical depth encountered by a light ray traversing the atmosphere at an impact parameter $b$ at wavelength $\lambda$, and integrates over all impact parameters (see Eq. 1-3 in \citealt{Ehrenreich2006} and figure \ref{fig:geometry} in this paper). By assuming hydrostatic equilibrium and spherical symmetry, the slant optical depth for the species $i$ along a chord at impact parameter $b$ is
\begin{equation}
\begin{split}
\label{optical_depth}
&\tau_{b,\,i}(\lambda) =X_i n_{0} \cdot \\
&\cdot \int_{-\infty}^{\infty}\sigma_{b,\,i}(\lambda, x)\dfrac{T_{0}}{T(\sqrt{b^{2}+x^{2}})}\exp \left[ -\int_{r_{0}}^{\sqrt{b^{2}+x^{2}}} \dfrac{dr}{H(r)}\right] dx\ ,
\end{split}
\end{equation}
where $n$ is the total particle numerical density in the atmosphere, $X_i = n_i/n$ is the volume mixing ratio of species $i$, $T$ is the temperature and $\sigma_{b,\,i}$ is the cross section profile along the considered chord. Quantities subscripted with 0 are taken at a reference level $r_{0}$ well inside the opaque part of the atmosphere of a gaseous giant or at the surface of a rocky planet. All of these quantities are functions of the position in the atmosphere. The typical local spatial vertical scale is given by the scale height defined by
\begin{equation}
H(r) = \dfrac{k_{\mathrm{B}}T(r)}{\mu(r) g(r)}\ ,
\end{equation}
where $g(r)$ is the gravity acceleration and $\mu(r)$ is the molecular weight of the species present in the atmosphere. If the atmosphere is composed of more than one species, the total optical depth is the sum
\begin{equation}
\label{eq:slant_optical depth}
\tau_{b}(\lambda) = \sum_i\tau_{b,\,i}(\lambda)\ .
\end{equation}
Integration of Eq. (\ref{eq:slant_optical depth}) over all the layers that contribute to a measurable atmospheric absorption gives a wavelength dependent atmospheric equivalent surface of absorption $\Sigma(\lambda)$, which is the surface of a completely optically thick disk that absorbs as much light as the entire translucent atmosphere.\\
A synthetic transmission spectrum is produced in three steps: computing the atmospheric structure (section \ref{sec:atm_profile}), computing the cross section and thus the optical depth of the absorbing species at each altitude layer (Sects. \ref{subsec:geometry}, \ref{subsec:c_sec}), and finally summing all the contributions from the layers.\\

\subsection{Atmospheric structure} \label{sec:atm_profile}
Hydrodynamical models of hot Jupiters show that the motion up to the lower part of the thermosphere is subsonic, which implies that hydrostatic balance is a good assumption (e.g. \citealt{Yelle2004}, \citealt{Koskinen2013}). In \pyeta the temperature profile can be assigned as an arbitrary function of the height in the atmosphere to keep generality. We introduce the possibility of assigning this quantity as a function of either $r$ or $p$ (pressure). Keeping $p$ as an independent variable allows us to account parametrically for temperature inversions, whose intensity is determined by the pressure in the atmosphere and not by the absolute height. The hydrostatic equilibrium equations are then integrated with the boundary condition that the pressure is $p_{0}$ at a reference layer $r_{0}$, both chosen by the user. If $p$ is the independent variable:
\begin{equation}
\label{hydrostatic height}
r\left (p\right ) =\left (\dfrac{1}{r_{0}} + \int_{p_{0}}^{p} \dfrac{k_{\mathrm{b}} T\left (p\right )}{m_{\mathrm{H}}\mu\left (p\right ) G M_{\mathrm{p}}p}dp\right )^{-1}\ .
\end{equation}
\\
\subsection{Geometry}
\begin{figure}
\def\svgwidth{\hsize}
\resizebox{\hsize}{!}{\includegraphics{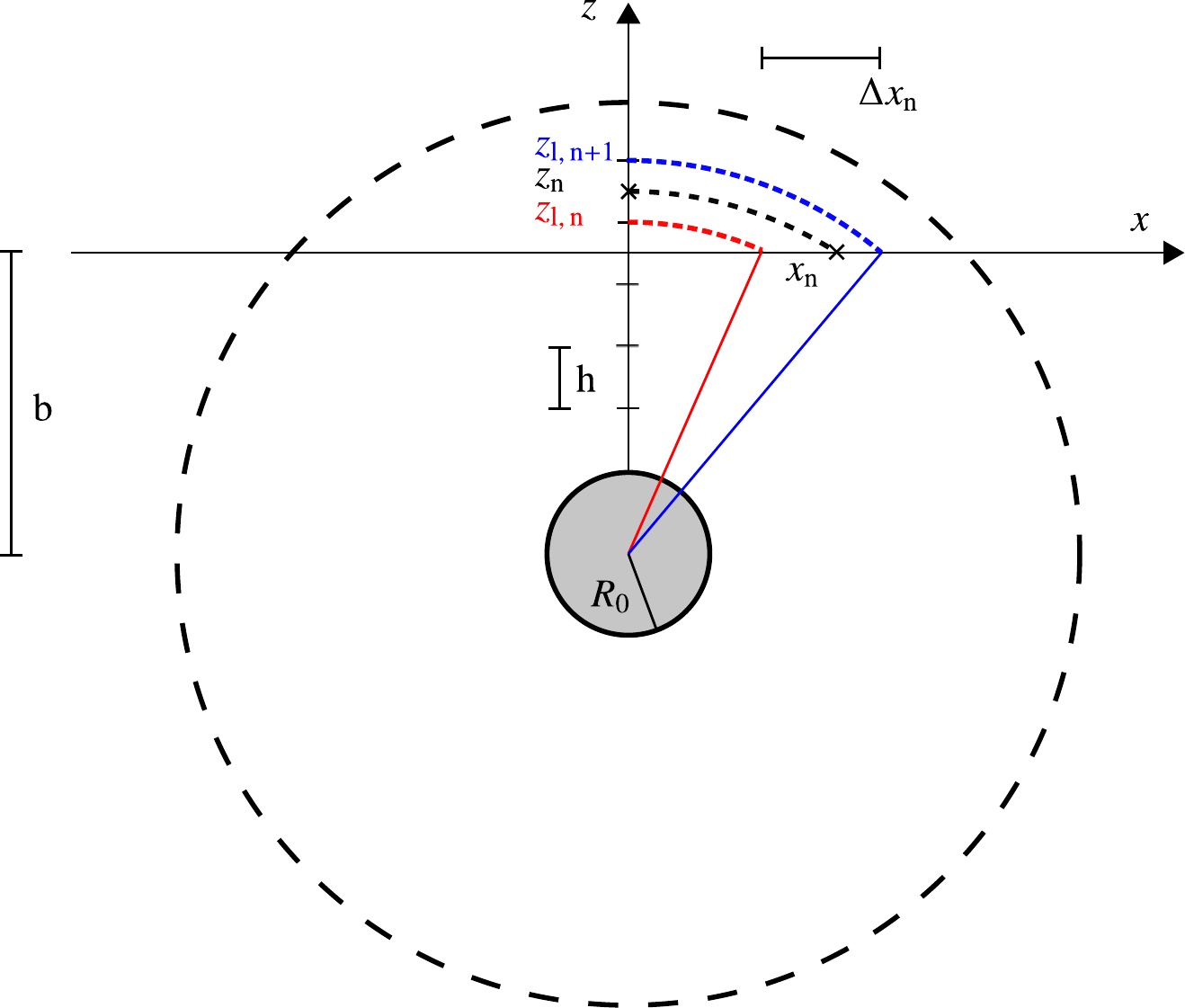}}
\caption{Geometry of transmission spectroscopy. \label{fig:geometry}}
\end{figure}

Because of the spectral resolution required for our models (millions of wavelength points in each layer), the implementation of geometry must compromise between time consumption and memory usage. With reference to Fig. (\ref{fig:geometry}), we divide the vertical axis $z$ in layers of height $h$. The optical depth in each layer $z_\mathrm{n}$ is computed and stored. We verified with a step-doubling method that an adaptive grid with four layers per local scale height is enough to grant accuracy in our case. The axis crossing the atmosphere ($x$) is divided into bins of width $\Delta x_\mathrm{n}$ which are the projection of $h$ along the chord. From geometric considerations: 
\begin{equation}
\Delta x_\mathrm{n} = \sqrt{\left(z_\mathrm{l,~n} + r_0\right)^2 - b^2} - \sqrt{\left(z_\mathrm{l,~n-1} + r_0\right)^2 - b^2}\ .
\end{equation}
The value of the optical depth in layer $z_\mathrm{n}$ can now be used as a proxy for the value of the optical depth in this bin. A typical run of the code then requires $\sim$ 3 GB of RAM and $\sim$ 10 minutes to run on an $\mathrm{Intel}^{\text{\textregistered}}$ $\mathrm{Xeon}^\text{\textregistered}$ CPU E5620 (2.40 GHz) using pre-computed opacity tables (see Sects. \ref{subsec:c_sec}).
\label{subsec:geometry}
\subsection{Photoabsorption cross sections at high spectral resolution} \label{subsec:c_sec}
Difficulties in accurately modelling the cross section $\sigma(\lambda)$ arise from poor theoretical knowledge of broadening, position and intensity of the lines and from the computational requirements set by the high number of lines to be considered in a line-by-line code.
\paragraph{Atomic lines:}
For bound-bound processes the cross section can be written as \citep{Seager2010, Ribicky1986}
\begin{equation}
\sigma_{\mathrm{at}}(\nu) = \dfrac{\pi e^{2}}{m_{e}c}f\Phi(\nu)\ ,
\end{equation}
where $e$ and $m_{e}$ are the elementary charge and mass of the electron, $c$ the speed of light, $f$ the oscillator strength of the transition and $\Phi(\nu)$ the line profile. Many physical effects determine the atomic line shape. Here we model each line as a Voigt profile, accounting for four types of broadening: thermal Doppler broadening, turbulent broadening, intrinsic broadening, and collisional broadening. We compute the Voigt profile as the real part of the Faddeeva function, computed with standard Python libraries (\texttt{scipy.special.wofz}).\\
The ASD line list \citep{nistspectra} provides the natural broadening widths, while for the collisional broadening of the sodium and potassium doublet, by far the dominant atomic transitions, we follow \cite{Iro2005} and \cite{Burrows2000} and set
\begin{equation}
\gamma_{\mathrm{coll},\,\left[\substack{\mathrm{Na} \\ \mathrm{K}}\right]} = \begin{bmatrix}0.071\\0.14\end{bmatrix}\left (\dfrac{T}{2000}\right )^{-0.7} \mathrm{cm}^{-1} \mathrm{atm}^{-1} \cdot P\cdot X_\mathrm{H_2}\ ,
\end{equation} 
where $X_\mathrm{H_2}$ is the volume mixing ratio of $H_{2}$, the relevant perturbing species.\\

\paragraph{Molecular transitions:}
We focus on water, which was detected in HD189733b using WFC3 data \citep{McCullough2014}. To model $\mathrm{H_{2}O}$ absorption, we adopt the HITEMP line list \citep{Rothman2013}\footnote{In appendix \ref{sec:appendix optimization} we discuss possible alternatives.}. Since the temperature in the atmosphere of HJs can reach well above 1000 K, it is necessary to use the HITEMP line list instead of the HITRAN line list \citep{Rothman2010}. The HITEMP line list includes transitions that may be of negligible intensity at laboratory temperature, but due to the temperature dependence of the line intensities might play a role at high temperatures \citep{Tinetti2007}. This results in the inclusion of $2.7\cdot 10^{7}$ $\mathrm{H_{2}O}$ transitions in the wavelength range $3\,300$--$20\,000\ \AA$ that interests us. Each line has to be modelled precisely at very high resolution ($\mathcal{R}\sim10^{6}$). The task requires state-of-the-art numerical techniques: we therefore rely on the HELIOS-K opacity calculator \citep{Grimm2015}. HELIOS-K is an ultrafast, open-source line-by-line opacity calculator for radiative transfer that extensively exploits parallel computing on GPUs to compute opacity tables generated from the HITRAN or HITEMP line list.\\
The shape of each line is modelled as a Voigt profile with a characteristic intensity, broadening and central wavelength. These are in turn determined by the thermodynamical conditions of the molecular gas:
\begin{itemize}
\item  A change in temperature has the effect of changing the partition function of a given molecule, i.e. the number density of the molecules associated with an energy level with respect to the total number density of that molecule. This impacts the intensity of a given line. 
\item A change in temperature or pressure changes the half width half maximum of a given line \citep{Mihalas1970, Rothman1995, Grimm2015, Hedges2016}.
\item A change in pressure induces a shift of the central wavelength of the line. 
\end{itemize}
All of these effects are considered in HELIOS-K, which is thus able to accurately model each line present in the HITEMP line list in the selected wavelength range.\\
We precomputed a grid of opacity tables covering the temperature range $500$--$3000~\mathrm{K}$ with a sampling of $50~\mathrm{K}$ and the pressure range $10^{-9}-10\ \mathrm{bar}$ in 21 logarithmically spaced points. Each opacity table is calculated on a wavenumber grid with a resolution of $0.01\ \mathrm{cm^{-1}}$ over the entire spectral range $3\,300$--$20\,000~ \AA$, corresponding to a wavelength resolution between $0.0033$ and $0.02~\mathrm{\AA}$ or $\mathcal{R} \gtrsim 5\cdot10^{5}$, which is enough to grant no opacity loss (see also \citealt{Hedges2016}). We extend the calculation for each line out to $25\ \mathrm{cm^{-1}}$ from its centre.\\
Our high-resolution approach is aimed at limiting molecular opacity loss. It is not optimal, however, since it requires a large amount of disk space to store the opacity tables and long computational times to model the entire spectral range. For the purposes of this paper, computational time is not an issue: it is sufficient to compute a limited number of models ($\sim10^2$, see section \ref{sec:methods}). In future implementations we will explore several possibilities to speed up the code, such as k-distribution tables \citep{Goody1989, Lacis1991, Fu1992, Grimm2015}, the opacity sampling method (\citealt{Sneden1976, Kitzmann2016}) and the recently proposed super-line approach \citep{Rey2016, Yurchenko2017}.

\paragraph{Continuum absorption}
\label{continuum}
The absorption lines emerge from a `continuum' absorption, due to Rayleigh scattering by several species and to scattering by aerosols when present.
We implement Rayleigh scattering due to $\mathrm{H_{2}}$. In \pyeta we adopt the prescription by \cite{Dalgarno1962, Dalgarno1965} commonly adopted in both solar system and exoplanets communities \citep{Seager2010}:
\begin{equation}
\label{eq:H2_rayleigh}
\sigma_{\mathrm{H_2}}(\lambda) = \dfrac{8.14\cdot10^{-13}}{\lambda^{4}} + \dfrac{1.28\cdot10^{-6}}{\lambda^{6}}+\dfrac{1.61}{\lambda^{8}}\ ,
\end{equation}
which is accurate to $O\left (\lambda^{-10}\right )$.
\begin{figure}
\resizebox{\hsize}{!}{\includegraphics{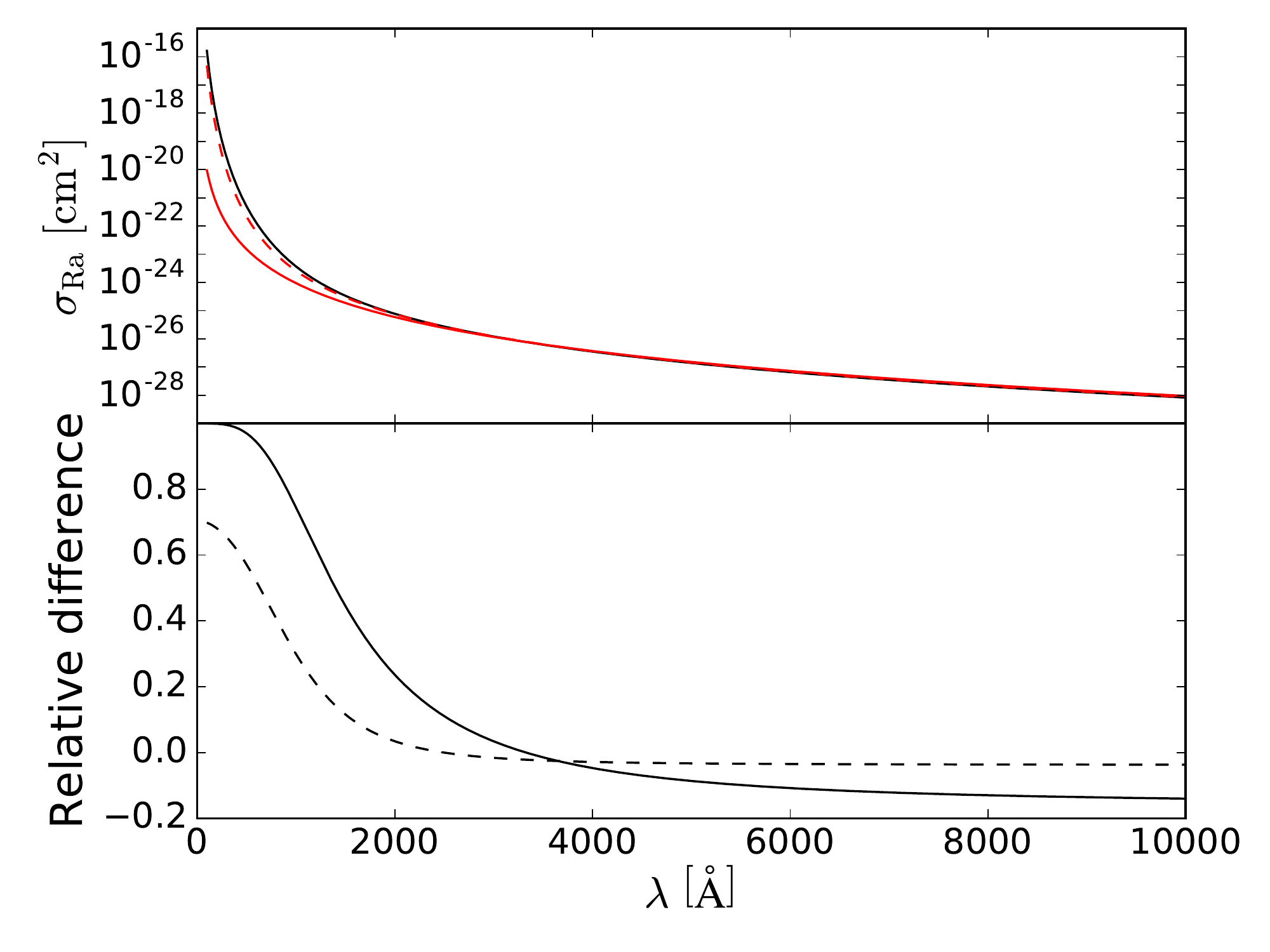}}
\caption{Comparison between different Rayleigh scattering prescriptions used in the literature. The black solid line is the prescription adopted in this work \citep{Dalgarno1962} and recommended by \citep{Seager2010}. The red solid line is the prescription adopted in the previous version of the program, based on \cite{Karplus1964}. Finally, the red dashed line is the version adopted by \cite{Heng2016}, based on \cite{Cox2000}. The relative difference between the prescriptions by \cite{Dalgarno1962} and \cite{Karplus1964} (lower panel, solid line) is negligible for wavelengths longer than $3\,300~\mathrm{\AA}$, of interest for this work. The difference between \cite{Dalgarno1962} and \cite{Cox2000} is negligible even bluewards. \label{fig:comparison_rayleigh}}
\end{figure}\\
\cite{Ehrenreich2006} implemented a different prescription for $\mathrm{H_{2}}$, based on the ab initio calculations given in \cite{Karplus1964}. A third alternative, based on \cite{Cox2000}, is adopted by other authors \citep{Heng2016}. In Fig. (\ref{fig:comparison_rayleigh}), we show a comparison between the Rayleigh cross sections for molecular hydrogen computed according to the prescriptions by \cite{Dalgarno1962}, \cite{Karplus1964}, and \cite{Cox2000}. They differ by $<10\%$ in the wavelength range $3\,300$--$20\,000~\AA$. We conclude that our analysis is consistent with all prescriptions. However, possible studies sensitive to Rayleigh scattering in the UV spectral range will have to consider which is the most appropriate law to follow.\\
We also include a simple treatment of aerosols as an opaque opacity source. Following \cite{Sing2016}, we distinguish between a grey absorber, to which we call `clouds', and a chromatic absorber that we generically call `hazes'\footnote{The naming convention we have adopted is commonly used by Earth scientists. The planetary science community generally distinguish clouds and hazes in terms of formation pathways rather than in terms of size. See also a blog entry by Sarah H\"orst on this topic:\\ \path{http://www.planetary.org/blogs/guest-blogs/2016/0324-clouds-and-haze-and-dust-oh-my.html}}. With this distinction, clouds mask the transmission spectrum below a pressure level $p_{\mathrm{c}}$ that coincides with the clouds top. Clouds are aerosols with particles that are larger than the wavelength of incident light. What we call hazes are, on the other hand, small particle aerosols (smaller than the wavelength of incident light) whose cross sections depend on wavelength and on the particle size. We model the global haze contribution as a scattering cross section:
\begin{equation}
\sigma_{\mathrm{h}} = \sigma_{\mathrm{h}}(\lambda_{0})\left (\dfrac{\lambda}{\lambda_{0}} \right )^{-s}\ .
\end{equation}
Here, $\sigma_{\mathrm{h}}(\lambda_{0}) = A\sigma_{\mathrm{H_2}}(\lambda_{0})$, where $A$ is the amplitude of the hazes scattering cross section in units of the $\mathrm{H_2}$ Rayleigh scattering cross section. $s$ is a slope that is determined by the composition and the sizes of the particles that scatter light \citep{Pont2013}. Observations are not able to distinguish all the different populations of scatterers (but see e.g. \citealt{Ehrenreich2014}); on the other hand leaving $s$ as a parameter allows us to have full generality when dealing with aerosol scattering.\\
In the limit for $\lambda\rightarrow\infty$, Eq. (\ref{eq:H2_rayleigh}) reduces to a pure power law with $s=4$ (Rayleigh scattering with a wavelength independent refraction index). In this case the cross sections of hazes and Rayleigh scattering by molecular hydrogen can be combined in an ``effective Rayleigh scattering'' cross section:
\begin{equation}
\label{eq:eff_rayleigh}
\sigma_{\mathrm{Ra,~eff}}(\lambda) = \sigma_{\mathrm{H_2}}(\lambda_{0})\left (\dfrac{\lambda}{\lambda_{0}} \right )^{-4}\left[ 1 + \mathrm{A} \left (\dfrac{\lambda}{\lambda_{0}} \right )^{4-s} \right]\ .
\end{equation}
From Eq. (\ref{eq:eff_rayleigh}), it is clear that hazes affect the transmission spectrum only if the scale parameter $\mathrm{A}$ is high enough. Depending on the index $s$ the effect may be more severe at longer wavelengths ($s < 4$) or shorter wavelengths ($s>4$).\\
Despite its simplicity, several studies based on similar prescriptions for clouds  and hazes showed that this treatment is sufficient to capture degeneracies between the presence of clouds and water abundance (e.g. \citealt{Kreidberg2015, Sing2016}).\\
In standard conditions, symmetric molecules such as $\mathrm{H}_{2}$ do not possess any dipole moment, and thus are not strong emitters. However, they may acquire a transient dipole during collisions with other molecules if the density is high enough. In extremely hydrogen-rich dense environments, such as cool low-metallicity stars \citep{Borysow1997, Morley2014} and some solar system planets \citep{Wordsworth2013, Zhang2015}, the absorption due to this phenomenon (collisional-induced absorption, CIA) should usually not be neglected. In our case of transmission geometry of a hot Jupiter, CIA contribution is negligible in the wavelength range considered (as molecular opacities are dominant at the pressures sounded). For consistency with the literature (e.g. \citealt{Madhusudhan2012, Moses2013, Malik2017}), we nonetheless include in \pyeta the CIA by $\mathrm{H}_{2}$--$\mathrm{H}_{2}$ and $\mathrm{H}_{2}$--$\mathrm{He}$ couples from the density-normalized HITRAN coefficients presented in \cite{Richard2012}.\\ 
Finally, we neglect the effects of refraction and multiple-scattering, working in approximation of pure absorption (where scattering opacity is treated as absorption opacity). \cite{Brown2001} and \cite{Burrows2000} claimed that these effects are unimportant in hot Jupiters. More recently, \cite{Betremieux2016a} argued that there may be a few cases in which this simplification is inaccurate (mainly cold and small planets, which we do not investigate here). \cite{Robinson2017} demonstrated that aerosol forward scattering can be an important consideration for planets whose host star is relatively large in angular size, as is the case for hot Jupiters.  However, for the models presented here, haze/cloud forward scattering would likely reduce our predicted transit depths by less than a scale height (\citealt{Robinson2017b}; Robinson, private communication).

\section{Simulating observed transmission spectra}
\label{compobs}
A comparison between theoretical models and observations is only possible after the following steps:
\begin{enumerate}
\item Convolution with the line spread function (LSF). Since the FWHM of the LSF varies by orders of magnitude between low- and high-resolution ($\sim0.048~\mathrm{\AA}$ for HARPS, up to $\sim~107.7\mathrm{\AA}$ for HST), its impact on the transmission spectrum in these different regimes must be taken into account;
\item Binning at the data sampling;
\item Only for ground-based data: wavelength normalization. Contrary to space-based observations, ground-based, high-resolution spectra are double differential (in time and wavelength, \citealt{Snellen2008, Wyttenbach2015, Heng2015}). The wavelength differentiation is required to remove time-varying signals due to the Earth's atmosphere or to the instrument. However, it removes the absolute reference for the absorption, a limitation that we need to take into account when making a comparison to models. 
\end{enumerate}

\subsection{Convolution with the instrumental LSF}
We model the LSF of all the instruments as Gaussians with full width half maximum (FWHM) equal to their nominal resolving power. The resolving power is defined as
\begin{equation}
\label{eq:resolving_power}
\mathfrak{R} = \dfrac{c}{\Delta v} = \dfrac{\lambda}{\Delta\lambda}\ ,
\end{equation}
where ${\Delta v}$ and ${\Delta\lambda}$ are the FWHM of one resolution element of the instrument considered in the velocity space and in the wavelength space, respectively. In Table \ref{tab:observations} we list the adopted parameters for the LSF of the instruments used for observations.\\
We compute the chromatic absorption due to the atmosphere of the planet, or transit depth, $\delta(\lambda)$. Then, we convolve this quantity with the LSF of the instrument,
\begin{equation}
\label{eq:convolution_delta}
\delta_{\mathrm{conv}}(\lambda) = (\delta \ast \mathrm{LSF})(\lambda) = \int_{-\infty}^{+\infty} \delta(\lambda)\,\mathrm{LSF}(\lambda - \lambda')~d\lambda'\ .
\end{equation}
To compute the integral on the right-hand term, it is necessary to limit its calculation in the wavelength space. We made sure that integration over 3 times the FWHM of the LSF causes no loss of flux for all simulated instruments.\\
\begin{table*}
\caption{Instrument set-ups of the observations used. For each data set, we indicate the wavelength range covered, the resolving power and the width of the instrumental LSF that we adopted. The HST data sets were also analysed by \cite{Sing2016}, who corrected the observations for the effect of stellar activity. We use their version of the HST transmission spectra in our work. \label{tab:observations}}
\centering
\begin{tabular}{lccccc}
\hline 
Instrument and set-up & $\lambda$ range [$\mathrm{\AA}$] & $\mathcal{R}$ & FWHM [km/s] & FWHM [$\mathrm{\AA}$] & Reference\\ 
\hline 
\\
HARPS & $3\,800$--$6\,900$  & $115\,000$ & 2.6 & $\sim0.048$ & \citealt{Wyttenbach2015}\\ 
HST STIS G430L  & $2\,900$--$5\,700$ & 530 & 565 & $\sim8$ & \citealt{Sing2011}\\ 
HST STIS G750M  & $5\,808$--$6\,380$ & $5\,000$  & 60 & $\sim1.2$ & \citealt{Huitson2012}\\ 
HST ACS HRC G800L & $5\,500$--$10\,500$ & 140  & $2\,140$ & $\sim40$ & \citealt{Pont2008}\\ 
HST WFC3 G141 & $11\,000$--$17\,000$ & 130  & $2\,300$ & $\sim108$ & \citealt{McCullough2014}\\ 
\\
\hline 
\end{tabular} 
\end{table*}
\subsection{Binning}
After the stellar light is dispersed, it hits the CCD pixels. In a calibrated spectrograph, each pixel is associated with a wavelength. The net effect is that the photons are binned by wavelength, each bin corresponding to a pixel. To compare the model with the observations it is necessary to bin the model in the same bins the observations are provided with.\\
Binning to the instrumental sampling must conserve flux, thus by
\begin{equation}
\label{average}
\delta_{\mathrm{bin}}\left( \lambda_i\right) = \dfrac{\int_{\lambda_i - dx_i}^{\lambda_i + dx_i} \delta_{\mathrm{conv}}(\lambda)~d\lambda}{2\,dx_i}~,
\end{equation}
where $\lambda_i$ and $dx_i$ represent the centre and the half width of a pixel in the wavelength space.
\subsection{Wavelength normalization}
To remove flux variations only due to the instrument or the atmosphere of Earth, ground-based observations require a wavelength differentiation \citep{Snellen2008, Wyttenbach2015, Heng2015}. Practically, the differential transmission spectrum $\mathfrak{\tilde{R}}(\lambda)$ presented in \cite{Wyttenbach2015, Wyttenbach2017} is normalized to a reference band. To compare the models to this data set, it is necessary to normalize the models in the same fashion. The relation between the transit depth and the differential transmission spectrum is
\begin{equation}
\label{eq:normalization}
\tilde{\mathfrak{R}}_{\mathrm{bin}}(\lambda_i) = \dfrac{1-\delta_{\mathrm{bin}}(\lambda_i)}{1-\delta_\mathrm{bin}(\lambda_\mathrm{ref})}\ ,
\end{equation}
where $\delta_{\mathrm{bin}}(\lambda_\mathrm{ref})$ is the transit depth after convolution with the LSF and binning, averaged in a reference band where no features are expected. Equation \ref{eq:normalization} is demonstrated in Appendix \ref{sec:Demonstration_normalization}.\\
\section{Transmission spectrum and planetary absorption}
\label{caveat}
\begin{figure*}
\centering%
\includegraphics[width=17cm]{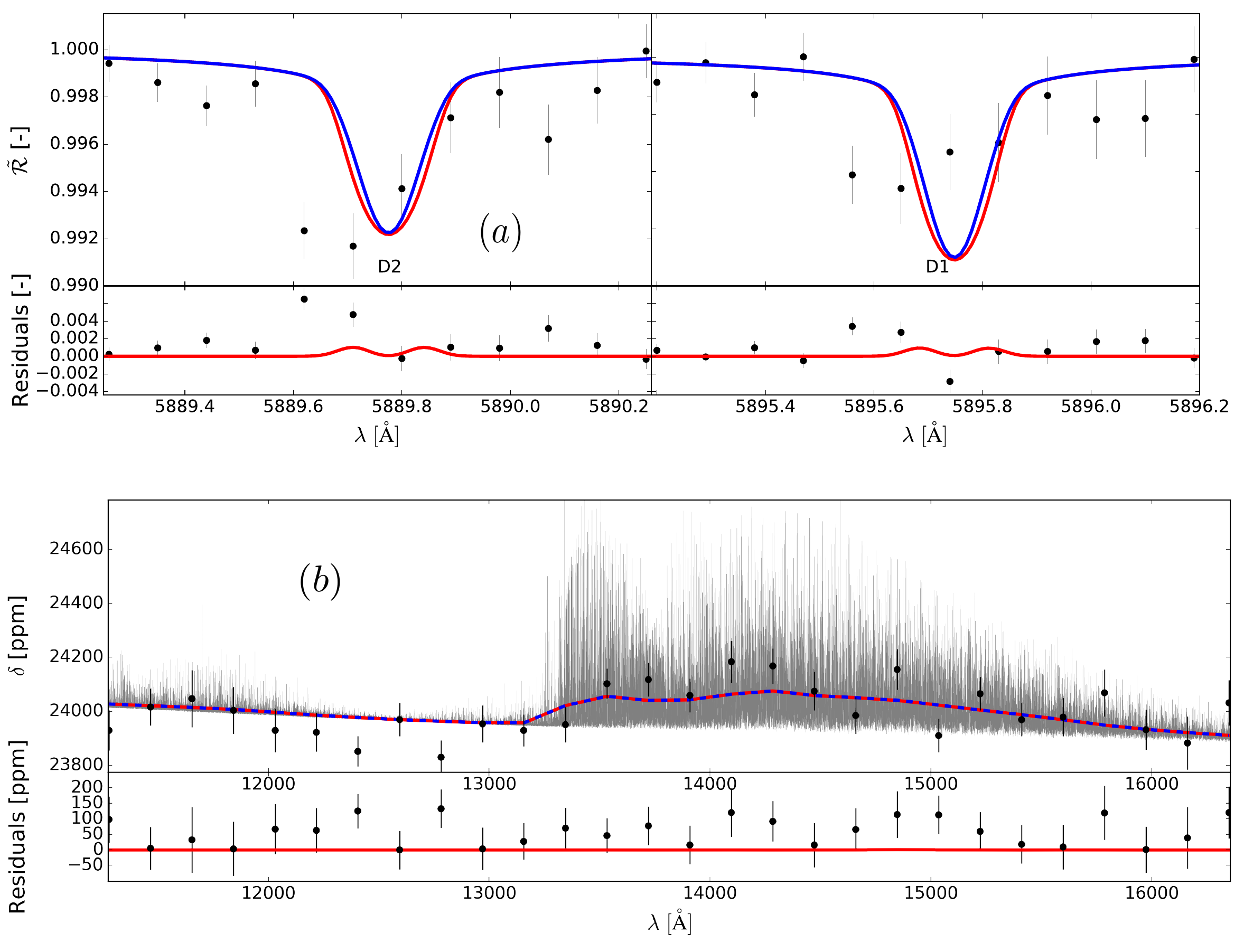}
\caption{Effect on the transmission spectrum of non-distributivity of convolution and binning with respect to division. In blue are shown models that are convolved and averaged in flux before division, which thus represent more closely the observed transmission spectrum; in red are shown models obtained by downgrading directly the planetary absorption, as usually done in the literature and throughout the rest of this paper. The black dots with error bars are the observed data. We also show the comparison of the residuals between data and the more accurate, blue model and the residuals between the models. Panel (a): In the case of HARPS data (binned 9x in the picture, but full resolution was used in the comparison), the difference in shape between the two models is visible by eye. However, the difference between the two models is still lower than the residuals between data and model (see lower panel). Panel (b): In the case of WFC3, the sharp short period variations (single molecular lines) are averaged out, while the longer period variations (molecular bands) are well sampled even at low-resolution. As a result, the difference between the models is also negligible. \label{Distributivity}}
\end{figure*}

The procedure we have outlined in section \ref{compobs} is commonly adopted to compare models and observations. However, it does not account for the fact that transmission spectra are built by dividing out stellar fluxes during the transit (in-transit, $F_{\mathrm{in}}$) and out-of-transit ($F_{\mathrm{out}}$). This is different from what is simulated by following the procedure outlined in the previous section, since neither convolution nor integration (or sum) are distributive with respect to division. For example, for convolution,
$$ \mathrm{LSF} \ast \left( f/g \right) \neq  \left(\mathrm{LSF} \ast f \right)/ \left(\mathrm{LSF} \ast g \right)~.$$
Thus, a more realistic simulation should compute as
\begin{equation}
\label{eq:true_obs_binning}
\delta_{\mathrm{bin}}\left( \lambda_i\right) = \dfrac{F_{\mathrm{in,\,bin}}\left( \lambda_i\right)}{F_{\mathrm{out,\,bin}}\left( \lambda_i\right)} = \dfrac{\int_{\lambda_i - dx_i}^{\lambda_i + dx_i} F_{\mathrm{in,~conv}}(\lambda)~d\lambda}{\int_{\lambda_i - dx_i}^{\lambda_i + dx_i} F_{\mathrm{out,~conv}}(\lambda)~d\lambda}~,\\
\end{equation}
\begin{equation}
\label{eq:true_obs_convolution}
F_{\mathrm{in/out,~conv}}(\lambda) = \int_{\sim3\cdot\mathrm{FWHM(LSF)}} F_{\mathrm{in/out}}(\lambda)\,\mathrm{LSF}(\lambda - \lambda')~d\lambda'\ .
\end{equation}
However, the computational cost of these more realistic simulations is doubled with respect to computing $\delta_{\mathrm{bin}}$ and $\tilde{\mathfrak{R}}_{\mathrm{bin}}$ directly from the theoretical planetary absorption $\delta(\lambda)$ (Eq. \ref{average}).\\
If the integrand can be approximated as a constant in the domain of integration it can pass through the sign of integral. If this applies to both Eq. (\ref{eq:true_obs_binning}) and Eq. (\ref{eq:true_obs_convolution}) the order of integration and division does not matter, and we can proceed following section  \ref{compobs}.\\
We test numerically if this is the case for our spectra. To perform the test, the stellar ($F_{\mathrm{out}}$) should be known. To represent it we use a PHOENIX model representative of the spectrum of HD189733 ($T=4900$~K, $\log g=4.5$, $\left[ \mathrm{Fe/H}\right]=0$, $\alpha=0$). The \pyeta code computes the absorption $\delta$ due to the planet, thus it is possible to compute $F_{\mathrm{in}} = F_{\mathrm{out}} (1-\delta)$. This is the stellar flux, planetary absorption included, before it is observed with an instrument. Then, the transmission spectra can be computed following Eq. (\ref{average}), Eq. (\ref{eq:true_obs_binning}), and Eq. (\ref{eq:true_obs_convolution}), and compared.\\
In Fig. (\ref{Distributivity}) we show the difference between the transmission spectrum computed using the stellar spectrum (Eq.\ref{eq:true_obs_binning} and Fig. \ref{Distributivity}, blue line), which is more accurate, and that using the absorption directly (Eq. \ref{average}, red line).
The statistical difference between the models is non-significant (using the BIC, defined in section \ref{sec:methods}, $\Delta$BIC $\sim$ 3 for HARPS, less than 1 for WFC3); this is also confirmed by the small residuals between the two models compared to the error bar (Fig. \ref{Distributivity}). We note, however, that for observations at higher signal-to-noise such as ESPRESSO, the E-ELTs, or JWST might deliver or with different host stars, this effect might become significant. Indeed, \cite{Deming2017} showed that neglecting this effect, which they call resolution-linked bias (RLB), yields a difference of 12\% in the peak transmittance for TRAPPIST-1b in the region around $1.4~\mu\mathrm{m}$. The effect is severe because cold stars likes TRAPPIST-1 have spectra dominated by forests of molecular lines. HD189733b is a K dwarf, thus the effect is less important.\\ 

\section{HD189733b case: methods}
\label{sec:methods}
We compare models of transmission spectra of HD189733b with optical, high-resolution differential transmission spectroscopy HARPS data (\citealt{Wyttenbach2015}) and optical, low-resolution HST STIS G430L and ACS HRC G800 (\citealt{Pont2008, Sing2011, Sing2016}), medium-resolution HST STIS G750M (\citealt{Huitson2012}) and NIR, low-resolution WFC3 (\citealt{McCullough2014}) data. The observations used are listed in Table \ref{tab:observations}. All the HST data are taken in the version presented by \citealt{Sing2016}, who corrects for possible systematic offsets between the data sets that are mainly due to stellar activity. For G750M, \cite{Sing2016} presents the data in wavelength bins of $60~\mathrm{\AA}$. However, a single point in the core of potassium is presented at higher sampling (bin width of $10~\mathrm{\AA}$). Its wavelength domain is already covered by the neighbouring lower sampling data points. We thus exclude this point from our analysis.\\
We calculate the $\chi^{2}$ between the HST transit depths or the normalized HARPS transmission spectrum in a wavelength bin $\lambda_i$, and the model processed as described in section \ref{compobs} in the same wavelength bin. We also compute the corresponding Bayesian information criterion (BIC, \citealt{Liddle2007}). The lowest BIC corresponds to the favoured model; a model whose BIC differs by more than 5 is statistically different. Moreover, the BIC takes into account the number of model parameters, preferring simpler models.
\subsection{HST data set}
We find a model that minimizes the $\mathrm{BIC_{HST}}$ for the full HST data set (best fit HST, BF-HST). Throughout this operation we fix the following parameters:
\begin{enumerate}
\item Star radius: $R_{\star} = 0.756\ R_{\odot}$;
\item Planetary mass: $M_{P}$ = $1.138\ {M_{\mathrm{J}}}$;
\item Constant solar composition, i.e. we adopt the following fixed mixing ratios: 
\begin{enumerate}
\item $X_{H_{2}} = 0.9289989$;
\item $X_{He} = 0.07$;
\item $X_{H_{2}O} = 10^{-3}$;
\item $X_{Na} = 10^{-6}$;
\item $X_{K}=10^{-7}$.
\end{enumerate}
\end{enumerate}
We adopt $p$ as an independent variable, and fix the $T-p$ profile. For the lower part of the atmosphere ($p>1~\mathrm{mbar}$), we adopt the profile presented in \cite{Sing2016}. For the thermosphere ($p<1~\mathrm{mbar}$), we adopt the profile presented in \cite{Wyttenbach2015}.\\
The remaining parameters are adjusted to minimize $\mathrm{BIC_{HST}}$. These parameters are:
\begin{itemize}
\item The reference radius, which we take as the radius at 10 bar: $r_{10\ \mathrm{atm}} = r(p=10\ \mathrm{atm})$ (but see \citealt{Heng2017} for a discussion on why this causes a degeneracy in inferring the values of mixing ratios at the order-of-magnitude level);
\item The hazes parameters $\sigma_\mathrm{h}(\lambda_0)$ and $s$;
\item The top pressure of the cloud deck $p_\mathrm{c}$.
\end{itemize}
The HST data set is further divided into optical (both STIS gratings and ACS) and WFC3. The optical data set is mainly sensitive to the haze parameters and to the reference radius, while the WFC3 data set is mainly sensitive to $p_\mathrm{c}$ and to the reference radius. Instead of performing a full retrieval, we exploit this difference to design a step-by-step optimization algorithm:
\begin{description}
\item[Step 1:] Adjust $r_{10\ \mathrm{atm}}$ using the average absorption in the WFC3 band;
\item[Step 2:] Adjust  $\sigma_\mathrm{h}(\lambda_0)$ and $s$ using STIS + ACS data;
\item[Step 3:] Adjust the cloud level $p_\mathrm{c}$ using the WFC3 data. For step 3, the full WFC3 spectral information is used. 
\end{description}
In each step, all parameters that are not being adjusted are fixed to values from the previous steps. If the parameters were perfectly uncorrelated, the procedure would converge to a minimum of $\mathrm{BIC_{HST}}$ after one iteration. This is not the case (e.g. optical and WFC3 data sets both depend on the reference radius), thus we iterate the three steps until convergence. Convergence is achieved when $\Delta\mathrm{BIC_{HST}}<5$ among the six models obtained in the last two iterations. The initialization for the first iteration assumes an aerosol-free atmosphere.\\
The intermediate nature between forward modelling and retrieval of our approach, which we refer to as `retrieval by hand', allows us to limit the number of models to be computed, and thus treat the problem numerically. At the same time, it allows us to reproduce the HST data set satisfactorily (see section \ref{results}).
\subsection{HARPS data set}
\begin{figure}
\centering%
\resizebox{\hsize}{!}{\includegraphics{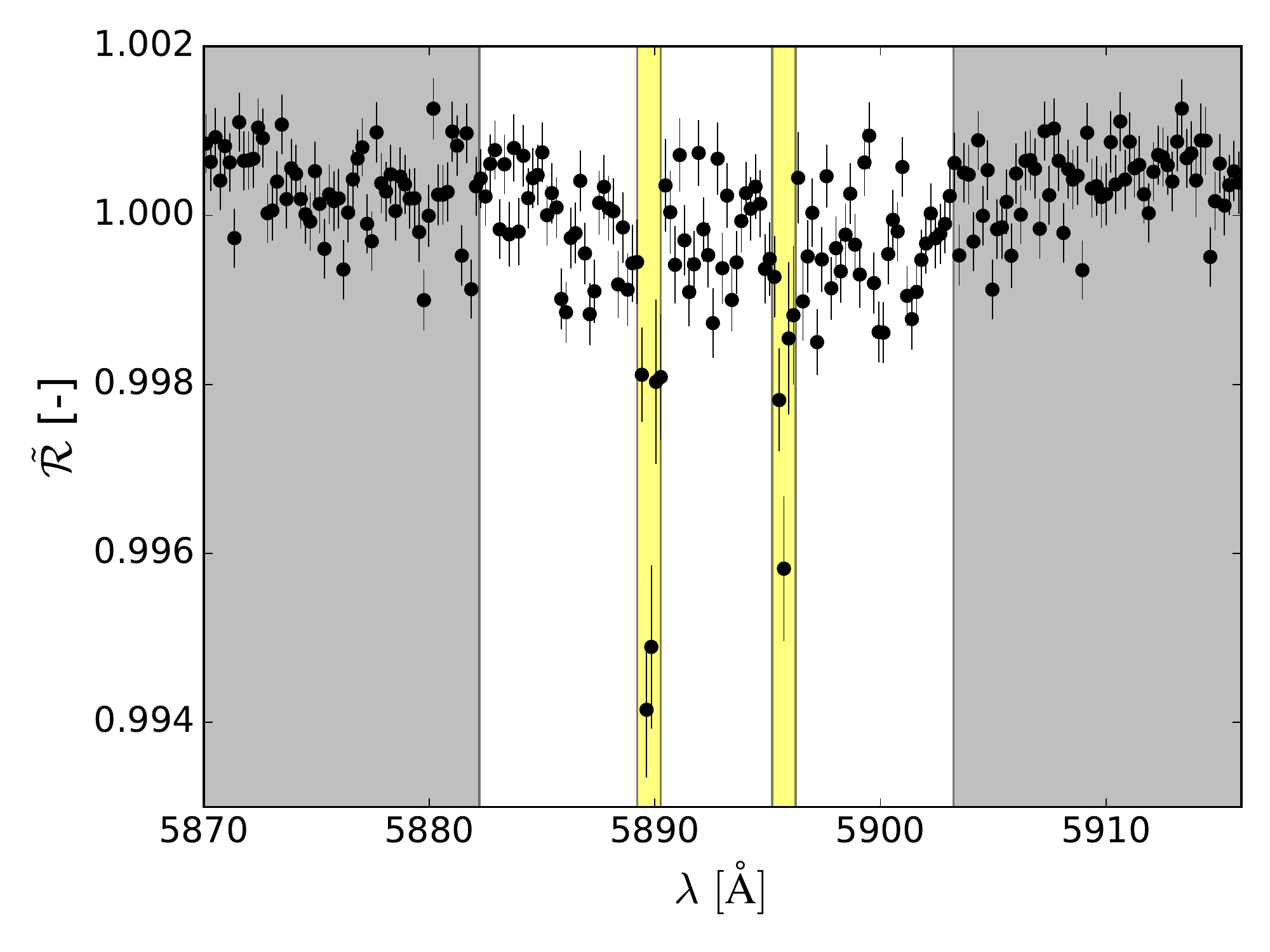}}
\caption{Bullet points represent the observed transmission spectrum previously published on \cite{Wyttenbach2015}, binned 21x (the full resolution data are used for the fit). The sodium excess absorption is clearly visible within the yellow bands where optimization is performed. In the grey bands no clear signal is found, and these are used to normalize the spectrum.\label{fig:HARPS_data_and_bands}}
\end{figure}

The BF-HST is then adjusted to reproduce the HARPS data set. For the wavelength normalization, we adopt the reference band $5870-5882.22~\mathrm{\AA},\ 5903.24-5916~\mathrm{\AA}$ (the same adopted by \citealt{Wyttenbach2015}, grey background in Fig. \ref{fig:HARPS_data_and_bands}). The merit functions are computed in two bands centred in the cores of the sodium lines: $5889.22~\mathrm{\AA}-5890.26~\mathrm{\AA}$ and $5895.20~\mathrm{\AA}-5896.24~\mathrm{\AA}$ (\citealt{Wyttenbach2015}, yellow background in  Fig. \ref{fig:HARPS_data_and_bands}). For each model, we compute the difference in BIC of the fit to the combined data set $\Delta\mathrm{BIC_{HR+HST}}$ with respect to the reference model BF-HST. We also compute the difference in BIC of the fit to the low- and high-resolution data sets separately ($\Delta\mathrm{BIC_{HR}}$ and $\Delta\mathrm{BIC_{HST}}$), to highlight which data set is driving the fit.\\
To adjust the HARPS data we modify a different set of parameters that mainly affects the thermosphere, leaving the troposphere mostly unchanged:
\begin{enumerate}
\item Thermospheric $T-p$ profile ($P<0.1~\mathrm{mbar}$);
\item Sodium abundance, $X_{Na}$.
\end{enumerate}
The strongest water lines may be affected by the $T-p$ profile modification and the sodium wings are produced in the troposphere. It is thus necessary to verify a posteriori that the quality of the fit to the HST data set is comparable to the quality obtained with the BF-HST for each modification.

\section{HD189733b case: results and discussion}
\label{results}

\subsection{HST data set}
\begin{figure*}
\centering
\includegraphics[width=17cm]{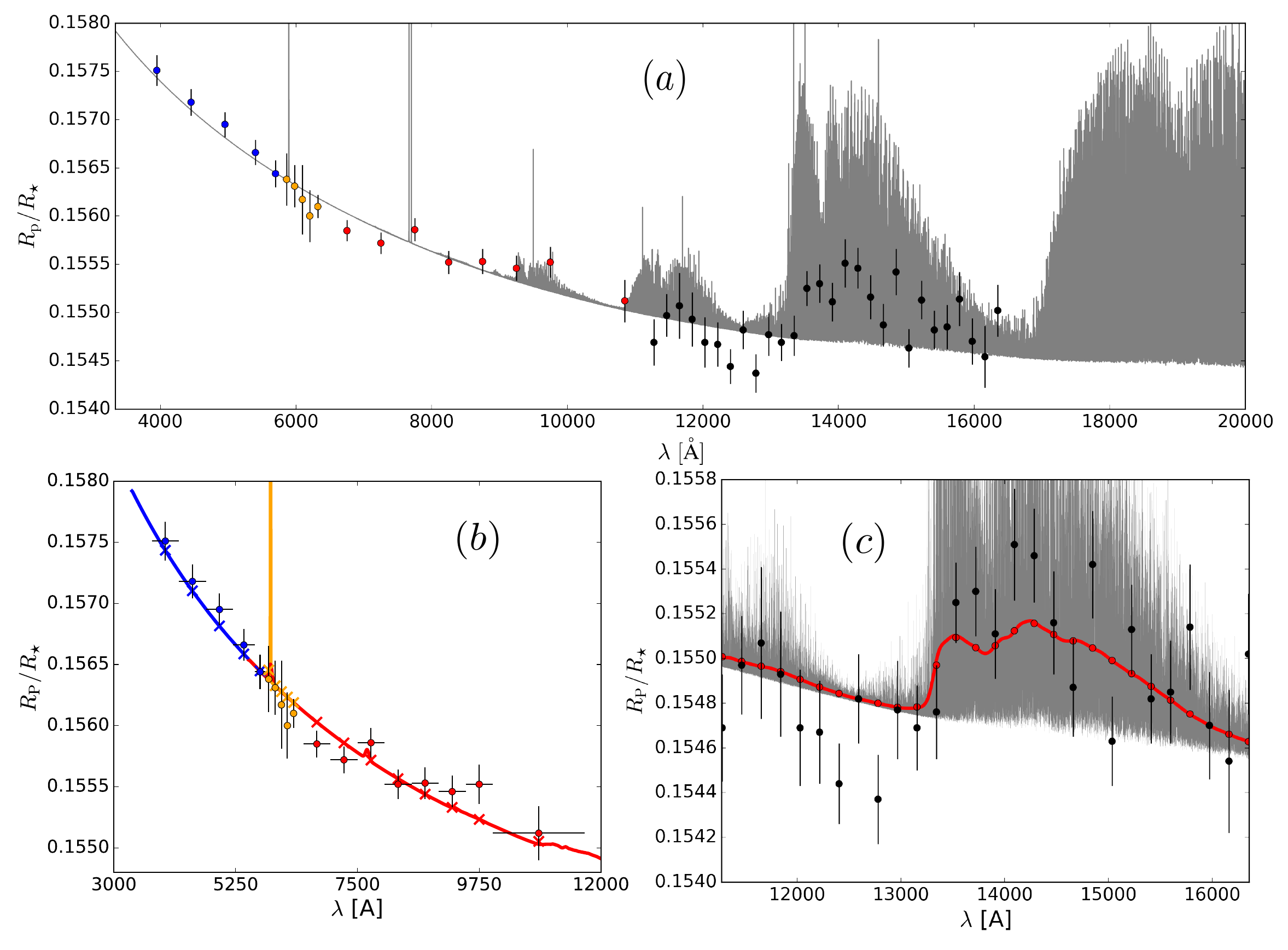}
\caption{BF-HST seen at various zoom levels. Panel (a): Whole transmission spectrum obtained after the parameters have been adjusted to reproduce the combined low-resolution data set. Panel (b): Zoom on the two STIS and the ACS data sets. The different colours correspond to models convolved with the different instrumental LSF (solid lines) and binned (dots). At the resolution of STIS G750M sodium is still quite sharp, while the lower resolution of ACS smears out the potassium doublet. This region of the spectrum is particularly sensitive to the presence of haze scattering. Panel (c): Zoom on the WFC3 data set. The water band is clearly visible; however, its intensity is reduced in comparison to an aerosol-free model. \label{fig:Best-fit}}
\end{figure*}

After six iterations we obtain a model that reproduces adequately the whole HST data set (BF-HST, see Fig. \ref{fig:Best-fit}). We find a reduced chi square $\chi_{\mathrm{HST,\,\nu=43}}^2 = 1.1$. When considering only the optical data sets (STIS G430L and G750M, and ACS), we obtain $\chi^2_{\nu=17} = 0.8$. For the WFC3 data set we find $\chi^2_{\nu=26} = 1.3$. The bluest points of the WFC3 observations would be reproduced better by the model if the scattering slope were weaker in this region, but a reduction of the haze content would cause a decrease in the quality of the reduced chi square obtained with the optical data set. The final model parameters are
\begin{itemize}
\item $\sigma_{\mathrm{h}} \sim8000\cdot \sigma_{\mathrm{H_2}}(589.46~\mathrm{nm})\left (\dfrac{\lambda}{589.46~\mathrm{nm}} \right )^{-3.17}~,$
\item $p_\mathrm{c} \sim 0.4~\mathrm{mbar}~,$
\item $r_{10\ \mathrm{atm}}\sim1.1082~\mathrm{R_{J}}~.$
\end{itemize}
Consistent with the literature, we find that the spectrum is dominated by aerosol scattering. An enhanced scattering due to a mix of different small particles (which produces an effective opacity with slope different than 4, see \citealt{Pont2013, Ehrenreich2014}) is required to explain the slope in the optical. The reduced intensity of the water feature probed by WFC3 is reproduced with a grey opacity cloud-deck.

\begin{figure*}
\centering
\includegraphics[width=17cm]{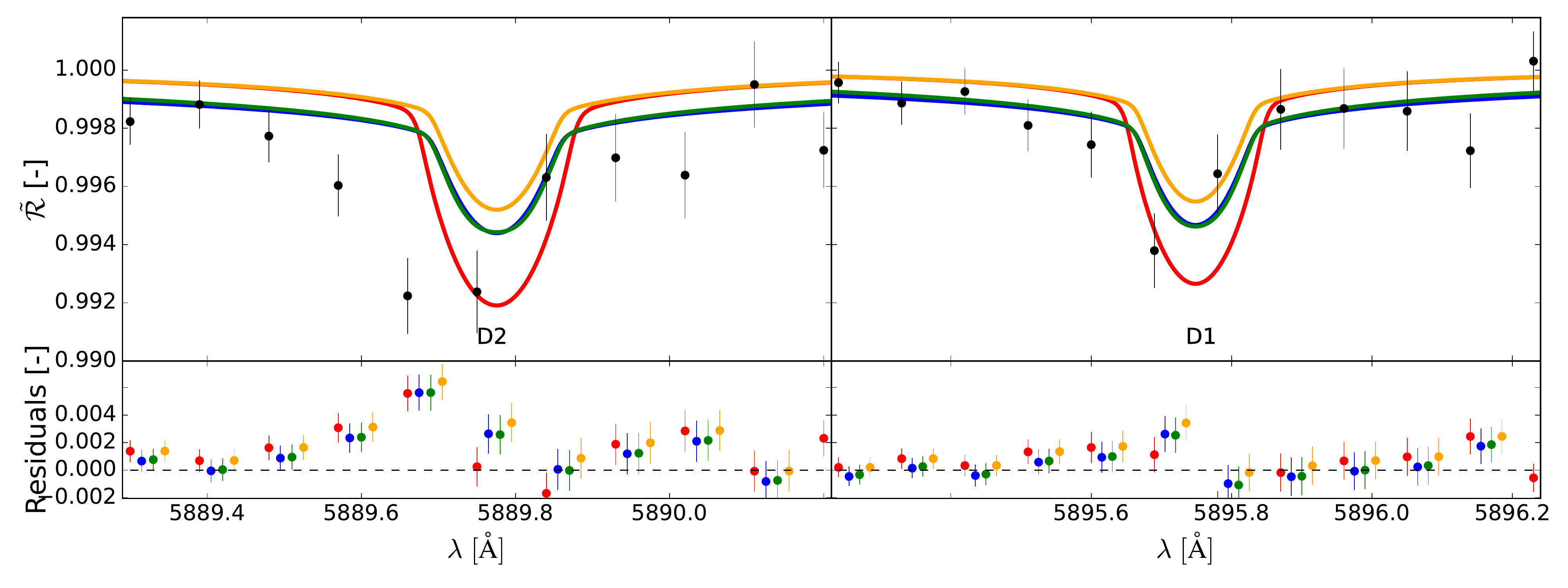}
\caption{Comparison between several models and the HARPS data set. Orange line: BF-HST. The sodium line cores absorption is too low to reproduce the high-resolution data set. Blue line: Aerosol-free model. This model reproduces better the high-resolution data set, but is not consistent with the low-resolution data set. Green line: 10Na. Our rigorous wavelength normalization process means it is indistinguishable from the aerosol-free model. It is, however, penalized compared to the previous model due to the presence of three additional parameters to describe aerosols. Red line: HTT. A higher thermospheric temperature increases the scale height of the outermost layers, and makes the sodium cores sharper. \label{fig:HARPS data}}
\end{figure*}

\begin{figure}
\resizebox{\hsize}{!}{\includegraphics{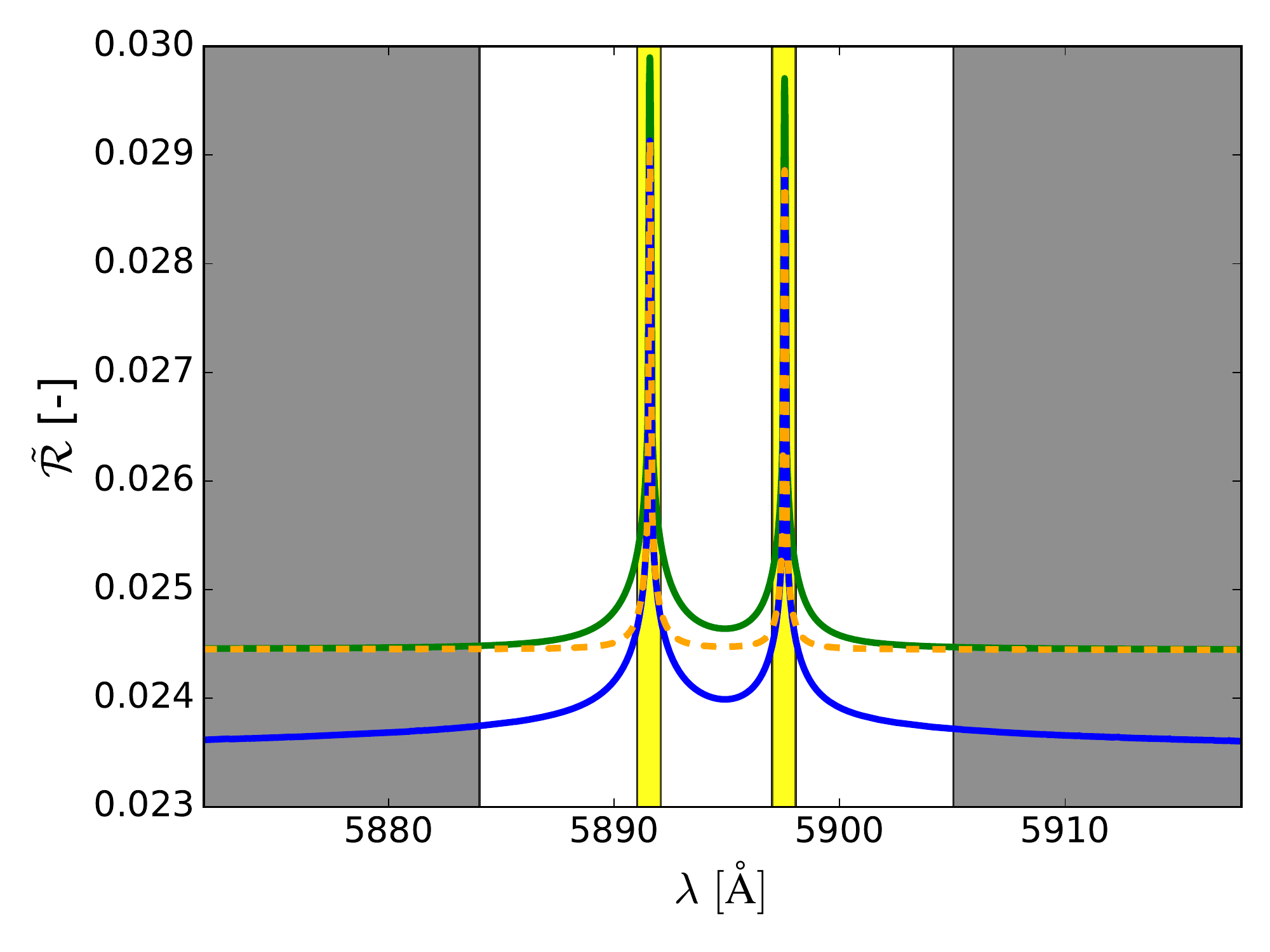}}
\caption{Theoretical models also shown in Fig. \ref{fig:HARPS data}, with the same colour-coding (orange: BF-HST, blue: AF, green: 10Na), but before wavelength normalization. The HTT model is not shown for clarity. Aerosols do not affect the cores of the sodium lines of the BF-HST model, which overlaps with the aerosol-free model in the yellow band used to compute the merit functions. However, they mask the wings in the reference bands (grey background). This example shows that high-resolution data are sensitive to the presence of aerosols. The 10Na model has a higher absorption in the sodium region than the aerosol-free model because the global optical depth is higher. However, HARPS alone cannot distinguish between the two cases (Fig. \ref{fig:HARPS data}) because of the double differential technique used to obtain transmission spectra. \label{fig:forward_model_comparison}}
\end{figure}

\begin{figure*}
\includegraphics[width=17cm]{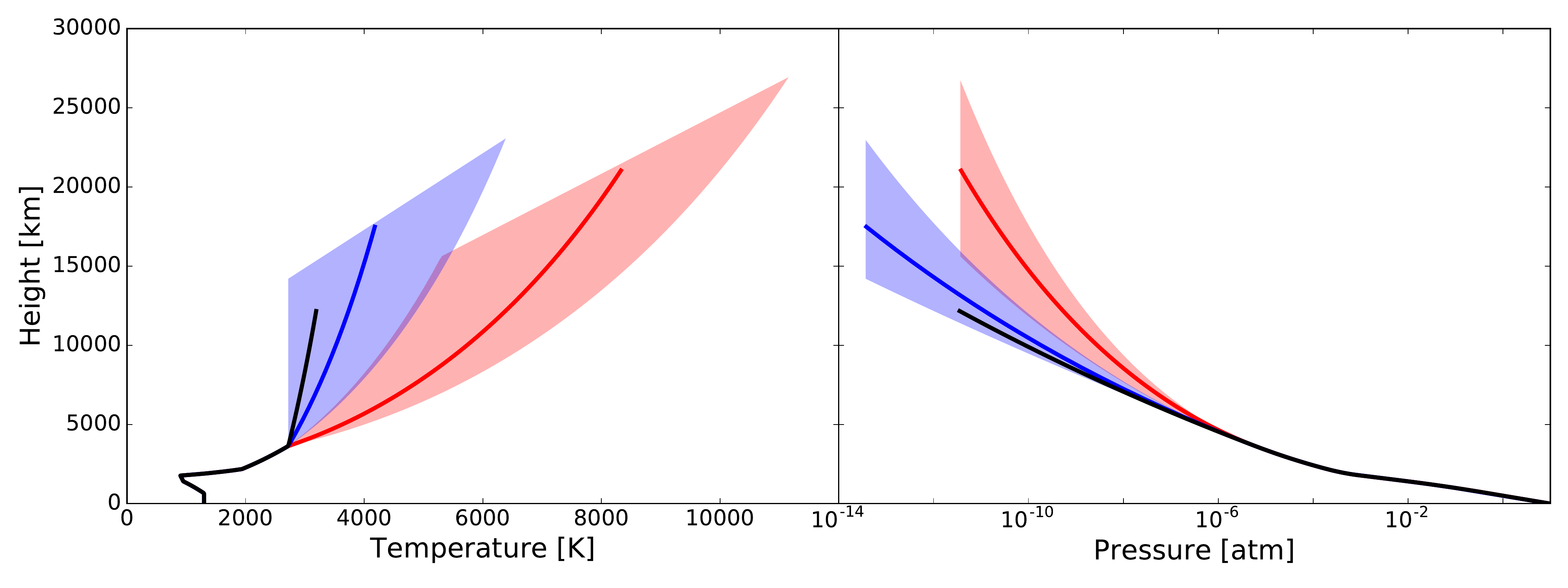}
\caption{Black solid line: $T-p$ profile adopted throughout the fit to the low-resolution data set. This is the combination of the temperature profile reported in \citealt{Sing2016} for the troposphere and by \citealt{Wyttenbach2015} for the thermosphere. Red solid line: Best choice to reproduce the high-resolution data set in the case of solar sodium abundance with aerosols at the $0.1~\mathrm{mbar}$ level (HTT). Red background: Region where variations in the $T-p$ profile produce models of similar quality to the HTT model ($\Delta\mathrm{BIC_{\mathrm{HR}}}<5$). The incompatibility between the HTT $T-p$ profile and the profile by \cite{Wyttenbach2015} is explained by the fact that they did not include aerosols in their analysis. Blue solid line and blue shadowed area: Same analysis for an assumed sodium abundance of 50 times the solar value. In this case the enhanced sodium abundance mimics an aerosol-free atmosphere such as that assumed by \cite{Wyttenbach2015}, thus the $T-p$ profiles are compatible. We did not allow for a negative temperature gradient in the $P<10^{-6}~\mathrm{bar}$ region, as it was not physically motivated.\label{TP-profile}}
\end{figure*}

\subsection{HARPS data set}
We show the results of the comparison with the HARPS data in Fig. \ref{fig:HARPS data}. The absorption in both sodium lines of the BF-HST model is low compared to the data (Fig. \ref{fig:HARPS data}). Since we adopted a $T-p$ profile consistent with \cite{Wyttenbach2015}, which in turn is a result of a fit to the same data set considered here, this comes as a partial surprise. The most likely reason is that \cite{Wyttenbach2015} assumed an aerosol-free atmosphere. Since the cores of the sodium lines are detected, aerosols are likely confined to the troposphere. The sodium core features probed by HARPS are generated higher up, in the lower thermosphere. However, it was already noticed from medium-resolution observations that scattering by aerosols may in some cases partially mask the sodium spectral feature (e.g. \citealt{Charbonneau2002, Pont2013, Sing2016, Heng2016}). To test whether aerosols in HD189733b are at high enough altitudes to have an effect on the high-resolution transmission spectrum we run an aerosol-free model (AF) that otherwise has the same parameters as the BF-HST.\\
The aerosol-free model reproduces the data better, as evinced by $\Delta\mathrm{BIC_{\mathrm{HR}}}\sim45$ and by smaller residuals. We can thus confirm that the discrepancy between our results and the results by \cite{Wyttenbach2015} is due to the presence of aerosols and their effect on wavelength normalization. Indeed, aerosols are not high enough to affect the inner cores of the lines, but they are high enough to occult the sodium wings in the reference bands (see Fig. \ref{fig:forward_model_comparison}). The difference in absorption between the core of the lines and the reference band is thus reduced in the presence of aerosols. Thanks to the correct wavelength normalization of the models, we are sensitive to this difference, and can distinguish between the BF-HST and its aerosol-free version.\\
However, the low-resolution data set firmly excludes the aerosol-free case; $\Delta\mathrm{BIC_{HST}}>10\,000$ supports the aerosol-rich case. To reconcile the space-borne and ground-based data sets we need to keep the aerosols and increase the contrast of the sodium feature. We thus modify the thermospheric $T-p$ profile to increase the scale height of the layers where the cores of the sodium lines are generated. This is obtained by increasing the thermospheric temperature with respect to the BF-HST. The high thermospheric temperature (HTT) model transmission spectrum is shown in Fig. \ref{fig:forward_model_comparison}. For this model we find $\Delta\mathrm{BIC_{HR}}\sim13$, indicating that it reproduces the HARPS data significantly better than the BF-HST. In Fig. \ref{TP-profile} we show the $T-p$ profile that best reproduces the HARPS data (red curve and shadowed area), and the comparison with \cite{Wyttenbach2015} (black curve). \\
Furthermore, the HTT model is also consistent with the HST data ($\Delta\mathrm{BIC_{\mathrm{HST}}}<1$), indicating that the troposphere is indeed left unchanged by the enhanced thermospheric heating. Finally, we note that our $T-p$ profile is in qualitative agreement with observation \citep{Vidal-Madjar2013} and models (e.g. \citealt{Yelle2004, Koskinen2013}) of the HD209458b thermosphere, with temperatures of up to $10\,000~\mathrm{K}$. HD209458b is in similar irradiation and gravity conditions to those of HD189733b, and both have expanded upper thermospheres.\\
The presence of aerosols also introduces a degeneracy between sodium abundance and thermospheric temperature. Intuitively, the more abundant the sodium is, the higher up in the atmosphere its features originate. For an abundance that is high enough, an aerosol-free atmosphere can be mimicked when focusing on the narrow wavelength band probed by high-resolution data. In Fig. \ref{fig:HARPS data}, we show in green a model with the same parameters as the BF-HST, but with 10 times its sodium abundance (10Na). The sodium lines are now generated higher up in the atmosphere, and part of their wings are generated above the aerosol scattering-deck (see Fig. \ref{fig:forward_model_comparison}, green line). As a result, when normalized, the enhanced sodium abundance model is nearly indistinguishable from the aerosol-free model in the $\chi^2$ band (see Fig. \ref{fig:HARPS data}). It also reproduces the HARPS data significantly better than the BF-HST. Models with 50 to 100 times the solar sodium abundance lead to even better matches with HARPS data. The quality of the fit to the HST data set is slightly decreased for such high sodium abundances because the wings of the doublets start to be visible. Overall, the BIC value resulting by the combined analysis of low- and high-resolution data favours the 50Na model.\\
The degeneracy between thermospheric heating and sodium abundance is further illustrated by constructing a second HTT model, with sodium abundance fixed at 50 times the solar value (HTT 50Na). Indeed, the difference in the global BIC between HTT 50Na and 50Na is not significant ($\Delta\mathrm{BIC_{\mathrm{HR+HST}}}\sim1.1$). In other words, a strongly (50x) enriched atmosphere would not require an enhanced thermospheric heating to exhibit the observed sodium signature (see also Fig. \ref{TP-profile}). We summarize our results in Table \ref{tab:BIC}.\\
We draw the following conclusions:
\begin{enumerate}
\item The low- to medium-resolution data set can be reconciled with the high-resolution data set;
\item The presence of aerosols at the $0.1~\mathrm{mbar}$ level, with which we explain the low- to medium-resolution data set, requires an enhanced thermospheric heating or enhanced sodium abundance in order to reproduce the high-resolution data set;
\item Having assumed solar sodium abundance, \cite{Wyttenbach2015} has underestimated the role of thermospheric heating because they assumed an aerosol-free atmosphere.
\end{enumerate}
Rather than the absolute abundance of sodium, it is its optical depth compared to scattering by aerosols that determines the contrast of its lines in the high-resolution transmission spectrum. This point is distinct but linked to the degeneracy between abundance and reference level first pointed out by \cite{Etangs2008} and generalized by \cite{Heng2017}. A complete analysis aimed at measuring sodium abundance or the precise value of the thermospheric temperature would thus need to fully explore this degeneracy, and is beyond the scope of this paper.\\
Finally, we note that the residuals between high-resolution data and models are asymmetric (see Fig. \ref{fig:HARPS data}). This could be due to the uncorrected Rossiter-McLaughlin effect \citep{Louden2015}. This effect has to be modelled before any quantitative conclusion can be obtained from the high-resolution data set. However, it does not affect our conclusion that high-resolution observations are sensitive to the set of parameters that complements lower resolution observations when the two techniques are combined.\\ 

\begin{table*}
\caption{BIC values computed for the models described in the text for the HST and HARPS data sets, and their combination. \label{tab:BIC}}
\centering
\begin{tabular}{lcccccccc}
\hline 
Model & $\sigma_{\mathrm{h}}(\lambda_0)$ & $s$ & $P_\mathrm{c}$~[mbar]& Thermospheric $T-p$ profile & $X_{Na}$ & $\Delta\mathrm{BIC_{HR}}$ & $\Delta\mathrm{BIC_{HST}}$ & $\Delta\mathrm{BIC_{HR+HST}}$\\ 
\hline 
\\
BF-HST & 8000 & -3.17 & 0.4 & \cite{Wyttenbach2015} & $10^{-6}$ & 0 & 0 & 0\\ 
AF & None & None & None & \cite{Wyttenbach2015} & $10^{-6}$  & -45.5 & $> 10^4$ & $> 10^4$\\ 
10Na & 8000 & -3.17 & 0.4 & \cite{Wyttenbach2015} & $10^{-5}$ & -29.1 & +0.5 & -28.6\\ 
50Na & 8000 & -3.17 & 0.4 & \cite{Wyttenbach2015} & $5\cdot10^{-4}$ & -37.7 & +2.2 & -35.5\\ 
100Na & 8000 & -3.17 & 0.4 & \cite{Wyttenbach2015} & $10^{-4}$ & -37.9 & +4.0 & -33.9\\ 
HTT & 8000 & -3.17 & 0.4 & Enhanced (red curve, Fig. \ref{TP-profile}) & $10^{-6}$ & -13.3 & -0.3 & -13.6\\ 
HTT 50Na & 8000 & -3.17 & 0.4 & Enhanced (blue curve, Fig. \ref{TP-profile}) & $5\cdot10^{-4}$ & -38.9 & +2.3 & -36.6\\
\\
\hline 
\end{tabular} 
\end{table*}

\section{Conclusions}
We developed a method to simultaneously compare theoretical models with high-resolution ($\mathcal{R}\sim10^5$), double-differential transmission spectra and low- to medium-resolution ($\mathcal{R}\sim10^2-10^3$) transmission spectra. It relies on a dedicated 1D, line-by-line, high resolution ($\mathcal{R}\sim10^6$) radiative transfer code, called $^\pi \eta$. Within this framework we have done the following:
\begin{itemize}
\item Showed that in the case of HD189733b, and likely many other planets, the common assumption that convolution and binning can be applied after dividing in-transit by out-of-transit spectra is justified by the error bars of current instrumentation (but see \citealt{Deming2017});
\item Built model atmospheres of the hot Jupiter HD189733b that are consistent with HST STIS, HST ACS, HST WFC3, and HARPS transmission spectra. These models are consistent with observations on an unprecedented range of pressures (tens of scale heights) from the troposphere to the thermosphere of the exoplanet. They show that the apparent discrepancy between flattened spectra observed at low-resolution and peaked features observed at high-resolution can be solved. This is obtained by adjusting a model that reproduces the low-resolution data to the high-resolution data, adopting a best set of parameters (sodium abundance, thermospheric $T-p$ profile) that affects the cores of the sodium lines while leaving the troposphere mainly unaffected;
\item Showed that ground-based, high-resolution observations are sensitive to the presence of aerosols. Neglecting aerosols has led to an underestimation of the thermospheric heating in this planet by \cite{Wyttenbach2015}. For a solar abundance of sodium, the same assumed by \cite{Wyttenbach2015}, we obtain thermospheric temperatures of up to $\sim10\,000~\mathrm{K}$, in qualitative agreement with models and observations of the similar hot Jupiter HD209458b;
\item Identified a degeneracy between sodium abundance and thermospheric heating that affects double-differential techniques in the presence of scattering by aerosols, linked to but distinct from the well-known degeneracy between abundance and reference level \citep{Etangs2008, Heng2017}. Retrievals of the absolute sodium abundance must take into account the presence of this degeneracy.
\end{itemize}
In summary, high-resolution ground-based observations provide unique insight into the thermosphere of exoplanets and are sensitive to alkali abundances when properly interpreted and combined with lower resolution observations. Our analysis opens new perspectives for the optimal exploitation of future facilities such as JWST on the one hand, and ESPRESSO and the E-ELTs on the other.

\begin{acknowledgements}
We thank R. Allart and the entire atmospheric group of the Geneva Observatory for insightful discussions and spot-on comments. We are also grateful to the anonymous referee, whose comments helped clarify several aspects of this paper, some of them crucial. This work has been carried out in the frame of the National Centre for Competence in Research ‘PlanetS’ supported by the Swiss National Science Foundation (SNSF). L.P., D.E., V.B., K.H., C.L., and F.P. acknowledge the financial support of the SNSF. This project has received funding from the European Research Council (ERC) under the European Union’s Horizon 2020 research and innovation programme (grant agreement No 724427).
\end{acknowledgements}

\bibliographystyle{aa}
\bibliography{\centralbibliography}

\begin{thebibliography}{97}
\expandafter\ifx\csname natexlab\endcsname\relax\def\natexlab#1{#1}\fi

\bibitem[{{Allart} {et~al.}(2017){Allart}, {Lovis}, {Pino}, {Wyttenbach},
  {Ehrenreich}, \& {Pepe}}]{Allart2017}
{Allart}, R., {Lovis}, C., {Pino}, L., {et~al.} 2017, \aap, 606, A144

\bibitem[{{Barber} {et~al.}(2006){Barber}, {Tennyson}, {Harris}, \&
  {Tolchenov}}]{Barber2006}
{Barber}, R.~J., {Tennyson}, J., {Harris}, G.~J., \& {Tolchenov}, R.~N. 2006,
  \mnras, 368, 1087

\bibitem[{{Barstow} {et~al.}(2017){Barstow}, {Aigrain}, {Irwin}, \&
  {Sing}}]{Barstow2017}
{Barstow}, J.~K., {Aigrain}, S., {Irwin}, P.~G.~J., \& {Sing}, D.~K. 2017,
  \apj, 834, 50

\bibitem[{{Ben-Jaffel} \& {Ballester}(2013)}]{Ben-Jaffel2013}
{Ben-Jaffel}, L. \& {Ballester}, G.~E. 2013, \aap, 553, A52

\bibitem[{{Benneke} \& {Seager}(2012)}]{Benneke2012}
{Benneke}, B. \& {Seager}, S. 2012, \apj, 753, 100

\bibitem[{{B{\'e}tr{\'e}mieux}(2016)}]{Betremieux2016a}
{B{\'e}tr{\'e}mieux}, Y. 2016, \mnras, 456, 4051

\bibitem[{{Birkby} {et~al.}(2017){Birkby}, {de Kok}, {Brogi}, {Schwarz}, \&
  {Snellen}}]{Birkby2017}
{Birkby}, J.~L., {de Kok}, R.~J., {Brogi}, M., {Schwarz}, H., \& {Snellen},
  I.~A.~G. 2017, \aj, 153, 138

\bibitem[{{Borysow} {et~al.}(1997){Borysow}, {Jorgensen}, \&
  {Zheng}}]{Borysow1997}
{Borysow}, A., {Jorgensen}, U.~G., \& {Zheng}, C. 1997, \aap, 324, 185

\bibitem[{{Bouchy} {et~al.}(2005){Bouchy}, {Udry}, {Mayor}, {Moutou}, {Pont},
  {Iribarne}, {da Silva}, {Ilovaisky}, {Queloz}, {Santos}, {S{\'e}gransan}, \&
  {Zucker}}]{Bouchy2005}
{Bouchy}, F., {Udry}, S., {Mayor}, M., {et~al.} 2005, \aap, 444, L15

\bibitem[{{Bourrier} {et~al.}(2013){Bourrier}, {Lecavelier des Etangs},
  {Dupuy}, {Ehrenreich}, {Vidal-Madjar}, {H{\'e}brard}, {Ballester},
  {D{\'e}sert}, {Ferlet}, {Sing}, \& {Wheatley}}]{Bourrier2013}
{Bourrier}, V., {Lecavelier des Etangs}, A., {Dupuy}, H., {et~al.} 2013, \aap,
  551, A63

\bibitem[{{Brogi} {et~al.}(2016){Brogi}, {de Kok}, {Albrecht}, {Snellen},
  {Birkby}, \& {Schwarz}}]{Brogi2016}
{Brogi}, M., {de Kok}, R.~J., {Albrecht}, S., {et~al.} 2016, \apj, 817, 106

\bibitem[{{Brogi} {et~al.}(2017){Brogi}, {Line}, {Bean}, {D{\'e}sert}, \&
  {Schwarz}}]{Brogi2017}
{Brogi}, M., {Line}, M., {Bean}, J., {D{\'e}sert}, J.-M., \& {Schwarz}, H.
  2017, \apjl, 839, L2

\bibitem[{{Brown}(2001)}]{Brown2001}
{Brown}, T.~M. 2001, \apj, 553, 1006

\bibitem[{{Burrows} {et~al.}(2000){Burrows}, {Marley}, \&
  {Sharp}}]{Burrows2000}
{Burrows}, A., {Marley}, M.~S., \& {Sharp}, C.~M. 2000, \apj, 531, 438

\bibitem[{{Charbonneau} {et~al.}(2002){Charbonneau}, {Brown}, {Noyes}, \&
  {Gilliland}}]{Charbonneau2002}
{Charbonneau}, D., {Brown}, T.~M., {Noyes}, R.~W., \& {Gilliland}, R.~L. 2002,
  \apj, 568, 377

\bibitem[{{Cox}(2000)}]{Cox2000}
{Cox}, A.~N. 2000, {Allen's astrophysical quantities}

\bibitem[{{Dalgarno} \& {Williams}(1962)}]{Dalgarno1962}
{Dalgarno}, A. \& {Williams}, D.~A. 1962, \apj, 136, 690

\bibitem[{{Dalgarno} \& {Williams}(1965)}]{Dalgarno1965}
{Dalgarno}, A. \& {Williams}, D.~A. 1965, Proceedings of the Physical Society,
  85, 685

\bibitem[{{de Kok} {et~al.}(2014){de Kok}, {Birkby}, {Brogi}, {Schwarz},
  {Albrecht}, {de Mooij}, \& {Snellen}}]{DeKok2014}
{de Kok}, R.~J., {Birkby}, J., {Brogi}, M., {et~al.} 2014, \aap, 561, A150

\bibitem[{{Deming} \& {Sheppard}(2017)}]{Deming2017}
{Deming}, D. \& {Sheppard}, K. 2017, \apjl, 841, L3

\bibitem[{{Ehrenreich} {et~al.}(2014){Ehrenreich}, {Bonfils}, {Lovis},
  {Delfosse}, {Forveille}, {Mayor}, {Neves}, {Santos}, {Udry}, \&
  {S{\'e}gransan}}]{Ehrenreich2014}
{Ehrenreich}, D., {Bonfils}, X., {Lovis}, C., {et~al.} 2014, \aap, 570, A89

\bibitem[{{Ehrenreich} {et~al.}(2006){Ehrenreich}, {Tinetti}, {Lecavelier Des
  Etangs}, {Vidal-Madjar}, \& {Selsis}}]{Ehrenreich2006}
{Ehrenreich}, D., {Tinetti}, G., {Lecavelier Des Etangs}, A., {Vidal-Madjar},
  A., \& {Selsis}, F. 2006, \aap, 448, 379

\bibitem[{{Ehrenreich} {et~al.}(2012){Ehrenreich}, {Vidal-Madjar}, {Widemann},
  {Gronoff}, {Tanga}, {Barth{\'e}lemy}, {Lilensten}, {Lecavelier Des Etangs},
  \& {Arnold}}]{Ehrenreich2012}
{Ehrenreich}, D., {Vidal-Madjar}, A., {Widemann}, T., {et~al.} 2012, \aap, 537,
  L2

\bibitem[{{Freedman} {et~al.}(2014){Freedman}, {Lustig-Yaeger}, {Fortney},
  {Lupu}, {Marley}, \& {Lodders}}]{Freedman2014}
{Freedman}, R.~S., {Lustig-Yaeger}, J., {Fortney}, J.~J., {et~al.} 2014, \apjs,
  214, 25

\bibitem[{{Fu} \& {Liou}(1992)}]{Fu1992}
{Fu}, Q. \& {Liou}, K.~N. 1992, Journal of Atmospheric Sciences, 49, 2139

\bibitem[{{Gandhi} \& {Madhusudhan}(2017)}]{Gandhi2017}
{Gandhi}, S. \& {Madhusudhan}, N. 2017, \mnras, 472, 2334

\bibitem[{{Goody} {et~al.}(1989){Goody}, {West}, {Chen}, \&
  {Crisp}}]{Goody1989}
{Goody}, R., {West}, R., {Chen}, L., \& {Crisp}, D. 1989, \jqsrt, 42, 539

\bibitem[{{Griffith}(2014)}]{Griffith2014}
{Griffith}, C.~A. 2014, Philosophical Transactions of the Royal Society of
  London Series A, 372, 20130086

\bibitem[{{Grimm} \& {Heng}(2015)}]{Grimm2015}
{Grimm}, S.~L. \& {Heng}, K. 2015, \apj, 808, 182

\bibitem[{{Hedges} \& {Madhusudhan}(2016)}]{Hedges2016}
{Hedges}, C. \& {Madhusudhan}, N. 2016, \mnras, 458, 1427

\bibitem[{{Heng}(2016)}]{Heng2016}
{Heng}, K. 2016, \apjl, 826, L16

\bibitem[{{Heng} \& {Kitzmann}(2017)}]{Heng2017}
{Heng}, K. \& {Kitzmann}, D. 2017, \mnras, 470, 2972

\bibitem[{{Heng} {et~al.}(2015){Heng}, {Wyttenbach}, {Lavie}, {Sing},
  {Ehrenreich}, \& {Lovis}}]{Heng2015}
{Heng}, K., {Wyttenbach}, A., {Lavie}, B., {et~al.} 2015, \apjl, 803, L9

\bibitem[{{Howe} \& {Burrows}(2012)}]{Howe2012}
{Howe}, A.~R. \& {Burrows}, A.~S. 2012, \apj, 756, 176

\bibitem[{{Huitson} {et~al.}(2012){Huitson}, {Sing}, {Vidal-Madjar},
  {Ballester}, {Lecavelier des Etangs}, {D{\'e}sert}, \& {Pont}}]{Huitson2012}
{Huitson}, C.~M., {Sing}, D.~K., {Vidal-Madjar}, A., {et~al.} 2012, \mnras,
  422, 2477

\bibitem[{{Iro} {et~al.}(2005){Iro}, {B{\'e}zard}, \& {Guillot}}]{Iro2005}
{Iro}, N., {B{\'e}zard}, B., \& {Guillot}, T. 2005, \aap, 436, 719

\bibitem[{{Karplus}(1964)}]{Karplus1964}
{Karplus}, M. 1964, \jcp, 41, 880

\bibitem[{{Kilpatrick} {et~al.}(2017){Kilpatrick}, {Cubillos}, {Stevenson},
  {Lewis}, {Wakeford}, {Macdonald}, {Madhusudhan}, {Blecic}, {Bruno},
  {Burrows}, {Deming}, {Heng}, {Line}, {Morley}, {Parmentier}, {Tucker},
  {Valenti}, {Waldmann}, {Bean}, {Beichman}, {Fraine}, {Krick}, {Lothringer},
  \& {Mandell}}]{kilpatrick2017}
{Kilpatrick}, B.~M., {Cubillos}, P.~E., {Stevenson}, K.~B., {et~al.} 2017,
  ArXiv e-prints [\eprint[arXiv]{1704.07421}]

\bibitem[{{Kitzmann}(2016)}]{Kitzmann2016}
{Kitzmann}, D. 2016, \apjl, 817, L18

\bibitem[{{Knutson} {et~al.}(2014){Knutson}, {Dragomir}, {Kreidberg},
  {Kempton}, {McCullough}, {Fortney}, {Bean}, {Gillon}, {Homeier}, \&
  {Howard}}]{Knutson2014}
{Knutson}, H.~A., {Dragomir}, D., {Kreidberg}, L., {et~al.} 2014, \apj, 794,
  155

\bibitem[{{Koskinen} {et~al.}(2013){Koskinen}, {Yelle}, {Harris}, \&
  {Lavvas}}]{Koskinen2013}
{Koskinen}, T.~T., {Yelle}, R.~V., {Harris}, M.~J., \& {Lavvas}, P. 2013,
  \icarus, 226, 1695

\bibitem[{Kramida {et~al.}(2015)Kramida, Ralchenko, Reader, \&
  Team}]{nistspectra}
Kramida, A.~E., Ralchenko, Y., Reader, J., \& Team, N. A. S.~D. 2015, {NIST
  Atomic Spectra Database (version 3.1.5)}

\bibitem[{{Kreidberg} {et~al.}(2015){Kreidberg}, {Line}, {Bean}, {Stevenson},
  {D{\'e}sert}, {Madhusudhan}, {Fortney}, {Barstow}, {Henry}, {Williamson}, \&
  {Showman}}]{Kreidberg2015}
{Kreidberg}, L., {Line}, M.~R., {Bean}, J.~L., {et~al.} 2015, \apj, 814, 66

\bibitem[{{Lacis} \& {Oinas}(1991)}]{Lacis1991}
{Lacis}, A.~A. \& {Oinas}, V. 1991, \jgr, 96, 9027

\bibitem[{{Lammer} {et~al.}(2003){Lammer}, {Selsis}, {Ribas}, {Guinan},
  {Bauer}, \& {Weiss}}]{Lammer2003}
{Lammer}, H., {Selsis}, F., {Ribas}, I., {et~al.} 2003, \apjl, 598, L121

\bibitem[{{Lavie} {et~al.}(2017){Lavie}, {Mendon{\c c}a}, {Mordasini}, {Malik},
  {Bonnefoy}, {Demory}, {Oreshenko}, {Grimm}, {Ehrenreich}, \&
  {Heng}}]{Lavie2016}
{Lavie}, B., {Mendon{\c c}a}, J.~M., {Mordasini}, C., {et~al.} 2017, \aj, 154,
  91

\bibitem[{{Lecavelier des Etangs} {et~al.}(2012){Lecavelier des Etangs},
  {Bourrier}, {Wheatley}, {Dupuy}, {Ehrenreich}, {Vidal-Madjar}, {H{\'e}brard},
  {Ballester}, {D{\'e}sert}, {Ferlet}, \& {Sing}}]{Etangs2012}
{Lecavelier des Etangs}, A., {Bourrier}, V., {Wheatley}, P.~J., {et~al.} 2012,
  \aap, 543, L4

\bibitem[{{Lecavelier Des Etangs} {et~al.}(2010){Lecavelier Des Etangs},
  {Ehrenreich}, {Vidal-Madjar}, {Ballester}, {D{\'e}sert}, {Ferlet},
  {H{\'e}brard}, {Sing}, {Tchakoumegni}, \& {Udry}}]{Etangs2010}
{Lecavelier Des Etangs}, A., {Ehrenreich}, D., {Vidal-Madjar}, A., {et~al.}
  2010, \aap, 514, A72

\bibitem[{{Lecavelier Des Etangs} {et~al.}(2008){Lecavelier Des Etangs},
  {Pont}, {Vidal-Madjar}, \& {Sing}}]{Etangs2008}
{Lecavelier Des Etangs}, A., {Pont}, F., {Vidal-Madjar}, A., \& {Sing}, D.
  2008, \aap, 481, L83

\bibitem[{{Liddle}(2007)}]{Liddle2007}
{Liddle}, A.~R. 2007, \mnras, 377, L74

\bibitem[{{Line} {et~al.}(2016){Line}, {Stevenson}, {Bean}, {Desert},
  {Fortney}, {Kreidberg}, {Madhusudhan}, {Showman}, \&
  {Diamond-Lowe}}]{Line2016}
{Line}, M.~R., {Stevenson}, K.~B., {Bean}, J., {et~al.} 2016, \aj, 152, 203

\bibitem[{{Louden} \& {Wheatley}(2015)}]{Louden2015}
{Louden}, T. \& {Wheatley}, P.~J. 2015, \apjl, 814, L24

\bibitem[{{MacDonald} \& {Madhusudhan}(2017)}]{MacDonald2017}
{MacDonald}, R.~J. \& {Madhusudhan}, N. 2017, \mnras, 469, 1979

\bibitem[{{Madhusudhan}(2012)}]{Madhusudhan2012}
{Madhusudhan}, N. 2012, \apj, 758, 36

\bibitem[{{Madhusudhan} {et~al.}(2014){Madhusudhan}, {Crouzet}, {McCullough},
  {Deming}, \& {Hedges}}]{Madhusudhan2014}
{Madhusudhan}, N., {Crouzet}, N., {McCullough}, P.~R., {Deming}, D., \&
  {Hedges}, C. 2014, \apjl, 791, L9

\bibitem[{{Madhusudhan} \& {Seager}(2009)}]{Madhusudhan2009}
{Madhusudhan}, N. \& {Seager}, S. 2009, \apj, 707, 24

\bibitem[{{Malik} {et~al.}(2017){Malik}, {Grosheintz}, {Mendon{\c c}a},
  {Grimm}, {Lavie}, {Kitzmann}, {Tsai}, {Burrows}, {Kreidberg}, {Bedell},
  {Bean}, {Stevenson}, \& {Heng}}]{Malik2017}
{Malik}, M., {Grosheintz}, L., {Mendon{\c c}a}, J.~M., {et~al.} 2017, \aj, 153,
  56

\bibitem[{{McCullough} {et~al.}(2014){McCullough}, {Crouzet}, {Deming}, \&
  {Madhusudhan}}]{McCullough2014}
{McCullough}, P.~R., {Crouzet}, N., {Deming}, D., \& {Madhusudhan}, N. 2014,
  \apj, 791, 55

\bibitem[{{Mihalas}(1970)}]{Mihalas1970}
{Mihalas}, D. 1970, {Stellar atmospheres}

\bibitem[{{Molli{\`e}re} {et~al.}(2017){Molli{\`e}re}, {van Boekel}, {Bouwman},
  {Henning}, {Lagage}, \& {Min}}]{Molliere2017}
{Molli{\`e}re}, P., {van Boekel}, R., {Bouwman}, J., {et~al.} 2017, \aap, 600,
  A10

\bibitem[{{Molli{\`e}re} {et~al.}(2015){Molli{\`e}re}, {van Boekel},
  {Dullemond}, {Henning}, \& {Mordasini}}]{Molliere2015}
{Molli{\`e}re}, P., {van Boekel}, R., {Dullemond}, C., {Henning}, T., \&
  {Mordasini}, C. 2015, \apj, 813, 47

\bibitem[{{Morley} {et~al.}(2014){Morley}, {Marley}, {Fortney}, {Lupu},
  {Saumon}, {Greene}, \& {Lodders}}]{Morley2014}
{Morley}, C.~V., {Marley}, M.~S., {Fortney}, J.~J., {et~al.} 2014, \apj, 787,
  78

\bibitem[{{Moses} {et~al.}(2013){Moses}, {Madhusudhan}, {Visscher}, \&
  {Freedman}}]{Moses2013}
{Moses}, J.~I., {Madhusudhan}, N., {Visscher}, C., \& {Freedman}, R.~S. 2013,
  \apj, 763, 25

\bibitem[{{Nascimbeni} {et~al.}(2013){Nascimbeni}, {Piotto}, {Pagano},
  {Scandariato}, {Sani}, \& {Fumana}}]{Nascimbeni2013}
{Nascimbeni}, V., {Piotto}, G., {Pagano}, I., {et~al.} 2013, \aap, 559, A32

\bibitem[{{Partridge} \& {Schwenke}(1997)}]{Partridge1997}
{Partridge}, H. \& {Schwenke}, D.~W. 1997, \jcp, 106, 4618

\bibitem[{{Pont} {et~al.}(2008){Pont}, {Knutson}, {Gilliland}, {Moutou}, \&
  {Charbonneau}}]{Pont2008}
{Pont}, F., {Knutson}, H., {Gilliland}, R.~L., {Moutou}, C., \& {Charbonneau},
  D. 2008, \mnras, 385, 109

\bibitem[{{Pont} {et~al.}(2013){Pont}, {Sing}, {Gibson}, {Aigrain}, {Henry}, \&
  {Husnoo}}]{Pont2013}
{Pont}, F., {Sing}, D.~K., {Gibson}, N.~P., {et~al.} 2013, \mnras, 432, 2917

\bibitem[{{Rey} {et~al.}(2016){Rey}, {Nikitin}, {Babikov}, \&
  {Tyuterev}}]{Rey2016}
{Rey}, M., {Nikitin}, A.~V., {Babikov}, Y.~L., \& {Tyuterev}, V.~G. 2016,
  Journal of Molecular Spectroscopy, 327, 138

\bibitem[{{Richard} {et~al.}(2012){Richard}, {Gordon}, {Rothman}, {Abel},
  {Frommhold}, {Gustafsson}, {Hartmann}, {Hermans}, {Lafferty}, {Orton},
  {Smith}, \& {Tran}}]{Richard2012}
{Richard}, C., {Gordon}, I.~E., {Rothman}, L.~S., {et~al.} 2012, \jqsrt, 113,
  1276

\bibitem[{{Robinson}(2017)}]{Robinson2017}
{Robinson}, T.~D. 2017, \apj, 836, 236

\bibitem[{{Robinson} {et~al.}(2017){Robinson}, {Fortney}, \&
  {Hubbard}}]{Robinson2017b}
{Robinson}, T.~D., {Fortney}, J.~J., \& {Hubbard}, W.~B. 2017, ArXiv e-prints
  [\eprint[arXiv]{1711.01278}]

\bibitem[{{Rothman} {et~al.}(2013){Rothman}, {Gordon}, {Babikov}, {Barbe},
  {Chris Benner}, {Bernath}, {Birk}, {Bizzocchi}, {Boudon}, {Brown},
  {Campargue}, {Chance}, {Cohen}, {Coudert}, {Devi}, {Drouin}, {Fayt}, {Flaud},
  {Gamache}, {Harrison}, {Hartmann}, {Hill}, {Hodges}, {Jacquemart}, {Jolly},
  {Lamouroux}, {Le Roy}, {Li}, {Long}, {Lyulin}, {Mackie}, {Massie},
  {Mikhailenko}, {M{\"u}ller}, {Naumenko}, {Nikitin}, {Orphal}, {Perevalov},
  {Perrin}, {Polovtseva}, {Richard}, {Smith}, {Starikova}, {Sung}, {Tashkun},
  {Tennyson}, {Toon}, {Tyuterev}, \& {Wagner}}]{Rothman2013}
{Rothman}, L.~S., {Gordon}, I.~E., {Babikov}, Y., {et~al.} 2013, \jqsrt, 130, 4

\bibitem[{{Rothman} {et~al.}(2010){Rothman}, {Gordon}, {Barber}, {Dothe},
  {Gamache}, {Goldman}, {Perevalov}, {Tashkun}, \& {Tennyson}}]{Rothman2010}
{Rothman}, L.~S., {Gordon}, I.~E., {Barber}, R.~J., {et~al.} 2010, \jqsrt, 111,
  2139

\bibitem[{{Rothman} {et~al.}(1995){Rothman}, {Wattson}, {Gamache}, {Schroeder},
  \& {McCann}}]{Rothman1995}
{Rothman}, L.~S., {Wattson}, R.~B., {Gamache}, R., {Schroeder}, J.~W., \&
  {McCann}, A. 1995, in \procspie, Vol. 2471, Atmospheric Propagation and
  Remote Sensing IV, ed. J.~C. {Dainty}, 105--111

\bibitem[{{Rybicki} \& {Lightman}(1986)}]{Ribicky1986}
{Rybicki}, G.~B. \& {Lightman}, A.~P. 1986, {Radiative Processes in
  Astrophysics}, 400

\bibitem[{{Seager}(2010)}]{Seager2010}
{Seager}, S. 2010, {Exoplanet Atmospheres: Physical Processes}

\bibitem[{{Seager} {et~al.}(2005){Seager}, {Richardson}, {Hansen}, {Menou},
  {Cho}, \& {Deming}}]{Seager2005}
{Seager}, S., {Richardson}, L.~J., {Hansen}, B.~M.~S., {et~al.} 2005, \apj,
  632, 1122

\bibitem[{{Seager} \& {Sasselov}(2000)}]{Seager2000}
{Seager}, S. \& {Sasselov}, D.~D. 2000, \apj, 537, 916

\bibitem[{{Sedaghati} {et~al.}(2015){Sedaghati}, {Boffin}, {Csizmadia},
  {Gibson}, {Kabath}, {Mallonn}, \& {Van den Ancker}}]{Sedaghati2015}
{Sedaghati}, E., {Boffin}, H.~M.~J., {Csizmadia}, S., {et~al.} 2015, \aap, 576,
  L11

\bibitem[{{Sing} {et~al.}(2016){Sing}, {Fortney}, {Nikolov}, {Wakeford},
  {Kataria}, {Evans}, {Aigrain}, {Ballester}, {Burrows}, {Deming},
  {D{\'e}sert}, {Gibson}, {Henry}, {Huitson}, {Knutson}, {Etangs}, {Pont},
  {Showman}, {Vidal-Madjar}, {Williamson}, \& {Wilson}}]{Sing2016}
{Sing}, D.~K., {Fortney}, J.~J., {Nikolov}, N., {et~al.} 2016, \nat, 529, 59

\bibitem[{{Sing} {et~al.}(2011){Sing}, {Pont}, {Aigrain}, {Charbonneau},
  {D{\'e}sert}, {Gibson}, {Gilliland}, {Hayek}, {Henry}, {Knutson}, {Lecavelier
  Des Etangs}, {Mazeh}, \& {Shporer}}]{Sing2011}
{Sing}, D.~K., {Pont}, F., {Aigrain}, S., {et~al.} 2011, \mnras, 416, 1443

\bibitem[{{Sing} {et~al.}(2015){Sing}, {Wakeford}, {Showman}, {Nikolov},
  {Fortney}, {Burrows}, {Ballester}, {Deming}, {Aigrain}, {D{\'e}sert},
  {Gibson}, {Henry}, {Knutson}, {Lecavelier des Etangs}, {Pont},
  {Vidal-Madjar}, {Williamson}, \& {Wilson}}]{Sing2015}
{Sing}, D.~K., {Wakeford}, H.~R., {Showman}, A.~P., {et~al.} 2015, \mnras, 446,
  2428

\bibitem[{{Sneden} {et~al.}(1976){Sneden}, {Johnson}, \& {Krupp}}]{Sneden1976}
{Sneden}, C., {Johnson}, H.~R., \& {Krupp}, B.~M. 1976, \apj, 204, 281

\bibitem[{{Snellen} {et~al.}(2008){Snellen}, {Albrecht}, {de Mooij}, \& {Le
  Poole}}]{Snellen2008}
{Snellen}, I.~A.~G., {Albrecht}, S., {de Mooij}, E.~J.~W., \& {Le Poole}, R.~S.
  2008, \aap, 487, 357

\bibitem[{{Snellen} {et~al.}(2010){Snellen}, {de Kok}, {de Mooij}, \&
  {Albrecht}}]{Snellen2010}
{Snellen}, I.~A.~G., {de Kok}, R.~J., {de Mooij}, E.~J.~W., \& {Albrecht}, S.
  2010, \nat, 465, 1049

\bibitem[{{Tennyson} \& {Yurchenko}(2012)}]{Tennyson2012}
{Tennyson}, J. \& {Yurchenko}, S.~N. 2012, \mnras, 425, 21

\bibitem[{{Tinetti} {et~al.}(2007){Tinetti}, {Liang}, {Vidal-Madjar},
  {Ehrenreich}, {Lecavelier des Etangs}, \& {Yung}}]{Tinetti2007}
{Tinetti}, G., {Liang}, M.-C., {Vidal-Madjar}, A., {et~al.} 2007, \apjl, 654,
  L99

\bibitem[{{Vidal-Madjar} {et~al.}(2013){Vidal-Madjar}, {Huitson}, {Bourrier},
  {D{\'e}sert}, {Ballester}, {Lecavelier des Etangs}, {Sing}, {Ehrenreich},
  {Ferlet}, {H{\'e}brard}, \& {McConnell}}]{Vidal-Madjar2013}
{Vidal-Madjar}, A., {Huitson}, C.~M., {Bourrier}, V., {et~al.} 2013, \aap, 560,
  A54

\bibitem[{{Vidal-Madjar} {et~al.}(2003){Vidal-Madjar}, {Lecavelier des Etangs},
  {D{\'e}sert}, {Ballester}, {Ferlet}, {H{\'e}brard}, \&
  {Mayor}}]{Vidal-Madjar2003}
{Vidal-Madjar}, A., {Lecavelier des Etangs}, A., {D{\'e}sert}, J.-M., {et~al.}
  2003, \nat, 422, 143

\bibitem[{{Vidal-Madjar} {et~al.}(2011{\natexlab{a}}){Vidal-Madjar}, {Sing},
  {Lecavelier Des Etangs}, {Ferlet}, {D{\'e}sert}, {H{\'e}brard}, {Boisse},
  {Ehrenreich}, \& {Moutou}}]{Vidal-Madjar2011}
{Vidal-Madjar}, A., {Sing}, D.~K., {Lecavelier Des Etangs}, A., {et~al.}
  2011{\natexlab{a}}, \aap, 527, A110

\bibitem[{{Vidal-Madjar} {et~al.}(2011{\natexlab{b}}){Vidal-Madjar}, {Sing},
  {Lecavelier Des Etangs}, {Ferlet}, {D{\'e}sert}, {H{\'e}brard}, {Boisse},
  {Ehrenreich}, \& {Moutou}}]{Vidal-Madjar2011b}
{Vidal-Madjar}, A., {Sing}, D.~K., {Lecavelier Des Etangs}, A., {et~al.}
  2011{\natexlab{b}}, \aap, 527, A110

\bibitem[{{Wordsworth}(2013)}]{Wordsworth2013}
{Wordsworth}, R. 2013, AGU Fall Meeting Abstracts

\bibitem[{{Wyttenbach} {et~al.}(2015){Wyttenbach}, {Ehrenreich}, {Lovis},
  {Udry}, \& {Pepe}}]{Wyttenbach2015}
{Wyttenbach}, A., {Ehrenreich}, D., {Lovis}, C., {Udry}, S., \& {Pepe}, F.
  2015, \aap, 577, A62

\bibitem[{{Wyttenbach} {et~al.}(2017){Wyttenbach}, {Lovis}, {Ehrenreich},
  {Bourrier}, {Pino}, {Allart}, {Astudillo-Defru}, {Cegla}, {Heng}, {Lavie},
  {Melo}, {Murgas}, {Santerne}, {S{\'e}gransan}, {Udry}, \&
  {Pepe}}]{Wyttenbach2017}
{Wyttenbach}, A., {Lovis}, C., {Ehrenreich}, D., {et~al.} 2017, \aap, 602, A36

\bibitem[{{Yelle}(2004)}]{Yelle2004}
{Yelle}, R.~V. 2004, \icarus, 170, 167

\bibitem[{{Yurchenko} {et~al.}(2017){Yurchenko}, {Amundsen}, {Tennyson}, \&
  {Waldmann}}]{Yurchenko2017}
{Yurchenko}, S.~N., {Amundsen}, D.~S., {Tennyson}, J., \& {Waldmann}, I.~P.
  2017, \aap, 605, A95

\bibitem[{{Zhang} {et~al.}(2015){Zhang}, {West}, {Irwin}, {Nixon}, \&
  {Yung}}]{Zhang2015}
{Zhang}, X., {West}, R.~A., {Irwin}, P.~G.~J., {Nixon}, C.~A., \& {Yung}, Y.~L.
  2015, Nature Communications, 6, 10231

\end{thebibliography}

\begin{appendix}
\section{Derivation of Eq. \ref{eq:normalization}}
\label{sec:Demonstration_normalization}
\paragraph{What is measured} The procedure followed by \cite{Wyttenbach2015, Wyttenbach2017} to obtain $\tilde{\mathfrak{R}}$ is the following:
\begin{enumerate}
\item All out-of-transit and in-transit spectra are normalized to a reference band:
\begin{equation}
\tilde{f}_{\mathrm{in/out}}(\lambda) = \dfrac{f_{\mathrm{in/out}}(\lambda)}{\left<f_{\mathrm{in/out}}(\lambda)\right>_{\lambda_{\mathrm{ref}}}}\ ,
\end{equation}
\item A master out $\tilde{F}_{\mathrm{out}}$ is built as the sum of the out-of-transit spectra and normalized;
\item The spectrum ratio is obtained as the sum of all the self-normalized in-transit spectra divided by the master out after having shifted them in the planetary rest frame:
\begin{equation}
\tilde{\mathfrak{R}}(\lambda)=\left.\sum_{\mathrm{transit}} \dfrac{\tilde{f_{\mathrm{in}}}(\lambda,\,t)}{\tilde{F}_{\mathrm{out}}(\lambda)}\right|_P\ .
\end{equation}
\end{enumerate}
\paragraph{What is simulated}
The quantity $\delta(\lambda)$ is the transit depth, i.e. the absorption due to the planet in units of stellar flux. This is then related to $f_\mathrm{in/out}$ by
\begin{equation}
\delta(\lambda) = \dfrac{f_{\mathrm{out}} - f_{\mathrm{in}}}{f_{\mathrm{out}}} = 1 - \dfrac{f_{\mathrm{in}}}{f_{\mathrm{out}}}\ ,
\end{equation}
and thus
\begin{equation}
\label{eq:delta_lambda_A}
\dfrac{f_{\mathrm{in}}}{f_{\mathrm{out}}} = 1 - \delta(\lambda)\ .
\end{equation}
Furthermore, our tests in section \ref{caveat} allow us to write
\begin{equation}
\label{eq:delta_lambda_B}
\dfrac{\left<f_{\mathrm{in}}\right>_{\lambda_{\mathrm{ref}}}}{\left<f_{\mathrm{out}}\right>_{\lambda_{\mathrm{ref}}}} \approx \left< \dfrac{f_{\mathrm{in}}}{f_{\mathrm{out}}} \right>_{\lambda_{\mathrm{ref}}} = 1 - \left< \delta(\lambda)\right>_{\lambda_{\mathrm{ref}}}\ .
\end{equation}
If we assume that the only time varying signal is due to the transit of the atmosphere of the planet, when we divide \ref{eq:delta_lambda_A} and \ref{eq:delta_lambda_B} we obtain Eq. (\ref{eq:normalization}).

\section{Choice and the use of the water line list}
\label{sec:appendix optimization}
The so-called `Million- to Billion-line radiative transfer challenge' \citep{Grimm2015} is a well known problem. The exact solution to the radiative transfer equation would require each single molecular absorption line to be modelled precisely (the `line-by-line' approach). For several molecules (e.g. water, methane) billions of lines contribute to the opacity at high temperatures ($>1\,000~\mathrm{K}$). To add complexity to the problem, multiple line lists are available. The impact of the choice of one line list over the others has never been fully quantified. We review here the knowledge for water to date.\\
The ExoMol project \citep{Tennyson2012} is the most up-to-date effort to provide line lists suitable for hot environments. In some cases this line list clearly outperforms previous efforts (e.g. for methane HITRAN vs ExoMol, \citealt{Lavie2016}). For water, however, there is general agreement that the line lists by \cite{Barber2006} and \cite{Partridge1997} are solid and complete, and indeed the BT2 line list by \cite{Barber2006} has been adopted by ExoMol (but a new line list is being developed inside the ExoMol project, Polyansky et al., in prep). The BT2 line list also constitutes the basis for the HITEMP water line list adopted in this paper. The HITEMP line list contains 25\% of the lines of the BT2 line list. The lines are selected to reduce the size of the line list while avoiding opacity losses at high-temperatures. This line list is adopted by several groups to model exoplanetary atmospheres (e.g. \citealt{Benneke2012, Molliere2015, Birkby2017, Molliere2017, MacDonald2017, Gandhi2017}). Furthermore, \cite{kilpatrick2017} provide a thorough analysis of the WFC3 spectrum of WASP-63b using forward modelling and retrieval methods based on both line lists. The authors conclude that there is `general agreement amongst all the results', suggesting that in this case differences due to the choice of the line list are minor. \cite{Brogi2017} pointed out that the HITEMP and the \cite{Freedman2014}\footnote{based on the BT2 line list} line lists may differ when dealing with high-resolution data; however, the culprit is not uniquely identified in their analysis (shift in line positions, line intensity, completeness, and different treatment of broadening are all possible candidates). It is thus not clear to what extent these differences would propagate to the comparison with low-resolution data. A quantitative assessment of the impact of different choices for the opacity tables is certainly warranted; because this assessment is lacking (to the best of our knowledge) we consider the HITEMP line list to be an appropriate choice to simulate WFC3 data.\\

\end{appendix}

\end{document}